\documentclass[12pt, draftclsnofoot,onecolumn]{IEEEtran} 
\usepackage{amsmath}
\usepackage{bm}
\usepackage{amssymb}
\usepackage{graphicx}
\usepackage{color, colortbl} 
\usepackage[dvipsnames,svgnames,x11names]{xcolor}
\usepackage{algpseudocode,algorithm}
\usepackage{hyperref}
\usepackage{enumerate}
\usepackage[caption=false,font=footnotesize]{subfig}

\newcommand{\splitatcommas}[1]{%
  \begingroup
  \ifnum\mathcode`,="8000
  \else
    \begingroup\lccode`~=`, \lowercase{\endgroup
      \edef~{\mathchar\the\mathcode`, \penalty0 \noexpand\hspace{0pt plus 1em}}%
    }\mathcode`,="8000
  \fi
  #1%
  \endgroup
}

\newcommand{\Ra}[1]{\textcolor{black}{#1}}
\newcommand{\Rb}[1]{\textcolor{black}{#1}}

\usepackage{cite}
\usepackage{url}
\usepackage{changes}
\begin{document}

\title{5G mmWave Cooperative Positioning and Mapping using Multi-Model PHD Filter and Map Fusion}
\author{
\IEEEauthorblockN{Hyowon Kim, \IEEEmembership{Student Member, IEEE}, Karl Granstr\"{o}m, \IEEEmembership{Member, IEEE}, Lin Gao, Giorgio Battistelli, Sunwoo Kim, \IEEEmembership{Senior Member, IEEE}, and Henk Wymeersch, \IEEEmembership{Senior Member, IEEE}}% <-this % stops a space
\thanks{H.~Kim and S.~Kim are with the Department of Electronics and Computer Engineering, Hanyang University, 04763 Seoul, South Korea (email: khw870511@hanyang.ac.kr; remero@hanyang.ac.kr).}
\thanks{K.~Granstr\"{o}m and H.~Wymeersch are with the Department of Electrical Engineering, Chalmers University of Technology, 412 58 Gothenburg, Sweden (email: henkw@chalmers.se; karl.granstrom@chalmers.se).}
\thanks{L.~Gao and G.~Battistelli are with the Dipartimento di Ingegneria dell'Informazione (DINFO), Universit\'{a} degli Studi di Firenze, 50139 Firenze, Italy (email: lin.gao@unifi.it; giorgio.battistelli@unifi.it).
}
% \thanks{This work was supported, in part, by the Swedish Research Council, under grant 2018-03701, and by the MSIT (Ministry of Science and ICT), Korea, under the ITRC (Information Technology Research Center) support program (IITP-2019-2017-0-01637) supervised by the IITP (Institute for Information \& Communications Technology Planning \& Evaluation).}
\thanks{This work has been accepted in the \textit{IEEE Transactions on Wireless Communications.}}
%for possible publication. Copyright may be transferred without notice, after which this version may no longer be accessible.}
}

\maketitle

\begin{abstract}
    5G millimeter wave (mmWave) signals can enable accurate positioning in vehicular networks when the base station and  vehicles are equipped with large antenna arrays. 
    However, radio-based positioning suffers from multipath signals generated by different types of objects in the physical environment. Multipath can be turned into a benefit, by building up a radio map (comprising the number of objects, object type, and object state) and using this map to exploit all available signal paths for positioning. %Building such a map is challenging, due to the inherent data association uncertainty, missed detection, and clutter. 
    We propose a new method for cooperative vehicle positioning and mapping of the radio environment, comprising 
    %in order to address these challenges. The proposed method comprises 
    a multiple-model probability hypothesis density filter and a map fusion routine, which is able to consider different types of objects and different fields of views. Simulation results demonstrate the performance of the proposed method. % handles the aforementioned challenges, and improves the vehicle positioning and mapping performance.
    %We propose a \textit{probability hypothesis density} (PHD) simultaneous localization and mapping (SLAM) filter and 
    %for estimating the vehicle state (location, heading, and clock bias) and mapping the radio environment in 5G vehicular networks.
    %In addition, the map obtained from the vehicle is periodically fused in the BS through the arithmetic average with our proposed fusion rules, subsequently the fused map 
    %Specifically, we 
    %Accurate positioning is promising due to 5G millimeter-wave (mmWave) signals from a single base station (BS) of which the transmitter and receiver is equipped with large antenna arrays.
    %We propose a \textit{probability hypothesis density} (PHD) simultaneous localization and mapping (SLAM) filter for estimating the vehicle state (location, heading, and clock bias) and mapping the radio environment in 5G vehicular networks.
    %In addition, the map obtained from the vehicle is periodically fused in the BS through the arithmetic average with our proposed fusion rules, subsequently the fused map 
    
\end{abstract}


\IEEEpeerreviewmaketitle

\section{Introduction}\label{sec:Introduction}
% 5G millimeter wave (mmWave) positioning is the next-generation cellular positioning framework, harnessing the potential of large bandwidths for accurate ranging, as well as large antenna arrays at a user and base station (BS) for accurate angle of arrival (AOA) and angle of departure (AOD) estimation \cite{WymSecDesDarTuf:J18}.
5G millimeter wave (mmWave) considers the potential of large bandwidths and large antenna arrays at the user and base station (BS), which enable accurate ranging, and angle of arrival (AOA) and angle of departure (AOD) estimation \cite{WymSecDesDarTuf:J18}.
Thus, 5G mmWave positioning is expected to be the next-generation cellular positioning framework \cite{Shahmansoori2018}.
Thanks to the aforementioned advantages of 5G mmWave, \Ra{measurements of} multipath components can be related to objects in the physical environment \cite{witrisal2016high,Shahmansoori2018,Palacios2019TWC}.
Therefore, it is possible to build up a map, which, e.g., can be reused by other users to cooperatively improve their position estimates.
%By relating the multipath components of 5G measurements to objects in the physical environment \cite{witrisal2016high,Shahmansoori2018,Palacios2019TWC}, it is possible to build a map, which, e.g., can be reused by other vehicles to {cooperatively} improve their position estimates.
Such maps can also remove the need for a priori synchronization of the users, and support single BS localization \cite{HenkGlobecom2018}.
% Hence, \textcolor{olive}{5G mmWave cooperative} positioning and mapping is a type of simultaneous localization and mapping (SLAM) problem \textcolor{olive}{\cite{Durrant2006SLAM1,Durrant2006SLAM2}.
% Mapping and positioning using 5G is termed 5G mmWave cooperative positioning and mapping, and can be categorized as a type of Simultaneous Localization and Mapping (SLAM) problem; refer to, e.g., \cite{Durrant2006SLAM1,Durrant2006SLAM2} for SLAM basics.
Mapping and positioning using 5G can be categorized as a simultaneous localization and mapping (SLAM) problem (for SLAM basics see, e.g., \cite{Durrant2006SLAM1,Durrant2006SLAM2}).
% \Karl{What do we want to say with the following sentence?}
In addition, 5G communication links can be used to share measurements, map, or location information, leading to cooperative  positioning and mapping. % can be considered by relaying the collected map information in the BS to each vehicle.
% by cooperation among vehicles without direct vehicle-to-vehicle communications, sharing of raw measurements, as well as performance relative measurements between vehicles.
% It is called as 5G mmWave cooperative positioning and mapping.
% In this paper, the map output by SLAM algorithm is represented by a set of static point landmarks,
% whose kinematic information is capsuled in the received multipath signals at the vehicle.
In 5G mmWave cooperative positioning and mapping \Ra{(i.e., positioning and mapping based on measurements from 5G mmWave communication signals)}, there are three main tasks: (i) \textit{Vehicle positioning:} determine the states (position, velocity, heading, clock bias) of the vehicles; (ii) \textit{Environment mapping:} estimate the number of objects, as well as each object's type and position; and (iii) \textit{Cooperation:} fusing the collected the map information from the vehicles, and relay it to each vehicle.
%\begin{enumerate}[(i)]
%    \item \textit{Vehicle positioning:} determine the states (position, velocity, heading, clock bias) of the vehicles; 
 %   \item \textit{Environment mapping:} estimate the number of objects, as well as each object's type and position; and 
  %  \item \EC{\textit{Cooperation:} fusing the collected the map information from the vehicles, and relay it to each vehicle.}
%\end{enumerate}
Even with the advantages of 5G mmWave, there are several challenges. First of all, due to the imperfect detection process at the receiver, there is a chance that objects that are inside the field-of-view (FoV) of vehicles are undetected. Secondly, the measurements may include false detections because of clutter, channel estimation error, and objects that are only visible in a short time. Third, since there are no origin-related tags on measurements, the data association must be be addressed, in either in an explicit or an implicit manner.
%To solve these general challenges of SLAM, a variety of approaches have been developed, see, e.g., \cite{yassin2018mosaic,aladsani2019leveraging,palacios2017jade,palacios2018communication,RicoTWC2018,HenkGlobecom2018,HyowonAsilomar2018,mahler_AES_2003_PHD,Mullane2011,Leung2013,BATTISTELLI20172457,Deusch2015,DeuschRD:2015,FatemiGSRH:2016_PMBradarmapping}, which will be detailed in Section \ref{sec:RelatedWorks} compared to this work.

To solve these general challenges of SLAM, a variety of approaches have been developed. They can be coarsely divided into three classes of methods (elaborated further in Section \ref{sec:RelatedWorks}): geometry-based \cite{yassin2018mosaic,aladsani2019leveraging,palacios2017jade,palacios2018communication}, based on message passing \cite{RicoTWC2018,HenkGlobecom2018,HyowonAsilomar2018}, and based on random finite sets (RFSs) \cite{mahler_AES_2003_PHD,Mullane2011,Leung2013,BATTISTELLI20172457,Deusch2015,DeuschRD:2015,FatemiGSRH:2016_PMBradarmapping}.
% in the context of this paper: \EC{5G SLAM (i.e., \textit{localization} and \textit{mapping} \cite{yassin2018mosaic,aladsani2019leveraging,palacios2017jade,palacios2018communication,RicoTWC2018,HenkGlobecom2018,HyowonAsilomar2018,mahler_AES_2003_PHD,Mullane2011,Leung2013,BATTISTELLI20172457,Deusch2015,DeuschRD:2015,FatemiGSRH:2016_PMBradarmapping,GranstromL:2013,GranstromRMS:2014,BlomBS1988,Mahler:2012,Pasha2009}) and map fusion (i.e., \textit{cooperation}) \cite{Li2018MSMOFOV,li2019distributed,BATTISTELLI20172457,li2017generalized,li2018local,Gao2019multiobject}, which will be detailed in Section \ref{sec:RelatedWorks} compared to this work.
Among RFS-based methods for tracking and mapping, {probability hypothesis density} (PHD) filters \cite{mahler_AES_2003_PHD} are widely used because they are computationally efficient alternatives that avoids explicit enumeration of the different data associations. Finally, in the \Ra{considered SLAM} problem, there are different types of measurements that are received by the vehicle, specifically measurements from the BS, scattering points, and reflecting surfaces \cite{HenkGlobecom2018,HyowonAsilomar2018}. These different types of measurements should be handled in the mapping using a multiple model approach, with one model for each measurement type. In addition, each vehicle has a limited FoV, and has thus only observed the parts of the map space that has been inside the FoV. It follows that the multi-object densities only contain information for a subset of the map space, specifically the parts that have been inside the FoVs. For this reason, a direct application of the standard fusion \cite{li2017generalized,li2018local,Gao2019multiobject} is not possible. %\Ra{First, the 5G radio environment considers multipath measurements generated by the different types of objects \cite{HenkGlobecom2018,HyowonAsilomar2018}, and different types of objects require different measurement models, one model for each type. 

In this paper, we address the aforementioned challenges and propose a new method for 5G mmWave cooperative positioning and mapping that is based on RFS theory. The proposed method comprises a Rao-Blackwellized (RB) representation of the joint vehicle-map density, with particle filters for the vehicle location and a multiple-model PHD filter for the map (i.e., PHD-SLAM), similar to \cite{Pasha2009}.
To update the particle weight, we theoretically derive a form of the set likelihood calculation. The PHD-SLAM filter is implemented by the Rao-Blackwellized particle filter (RBPF).
Cooperation is handled using asynchronous map fusion through a modified arithmetic average (AA) approach, taking into account the different FoVs of the vehicles by designing fusion weights for AA map fusion by first decomposing the map space into mutually disjoint subsets, and then selecting weights for each subset.
The BS performs map fusion with one vehicle at a time, so that through multiple interactions, each vehicle contributes to, and has access to, the global BS map.
%\Ra{To handle the first challenge, we develop a multiple-model PHD-SLAM filter similar to the on presented in \cite{Pasha2009}, that allows joint estimation of both the object types and the object states.}
%
%For each vehicle particle, the MM-PHD filter jointly estimates the objects' locations and types. 
%
% \Ra{Although we adopt the PHD filter \cite{mahler_AES_2003_PHD} for computationally efficient mapping and the AA method for map fusion, there are additional challenges.}
%
%\Ra{To handle the first challenge, we develop a multiple-model PHD-SLAM filter similar to the on presented in \cite{Pasha2009}, that allows joint estimation of both the object types and the object states.
%
%The details of differences of this work compared to the previous works will be discussed in Sec. \ref{sec:RelatedWorks}.
%
% We can coarsely categorize \EC{5G SLAM (i.e., \it{localization} and \it{mapping})} into three classes: methods based on geometry \cite{yassin2018mosaic,aladsani2019leveraging,palacios2017jade,palacios2018communication}, methods based on message passing \cite{RicoTWC2018,HenkGlobecom2018,HyowonAsilomar2018}, and methods based on random finite set (RFS) theory.
% \EC{Map fusion works for \textit{cooperation} are also categorized into two frameworks: methods based on generalized covariance intersection (GCI) \cite{Li2018MSMOFOV,li2019distributed,BATTISTELLI20172457}, and methods based on arithmetic average (AA) \cite{li2017generalized,li2018local,Gao2019multiobject}.
% The above previous works will be detailed in Section \ref{sec:RelatedWorks}.
% }
%
The main contributions of the paper are summarized as follows:
    \begin{itemize}
        \item For the \Ra{propagation environment} with multiple objects, multiple object types, and multipath measurements, we present and evaluate a novel solution to the cooperative positioning and mapping problem, based on a RB representation of the joint position-map density and a multiple-model PHD representation of the map, as well as a novel AA fusion rule. 
        %\item With focusing on the 5G scenario, the derived 5G mmWave cooperative positioning and mapping can deal with different types of objects generating multipath measurements, based on multiple-model PHD filter.
        % \item The derived 5G mmWave cooperative positioning and mapping can deal with different types of objects generating multipath measurements, based on multiple-model PHD filter.
        % \item We fuse information from different vehicles at a fusion center though asynchronous map fusion, thus significantly speeding up the mapping process.
        %\item We propose a map fusion method that makes it possible for vehicles to reuse the map built by other vehicles, thus significantly speeding up the mapping the propagation environment.
        \item We derive a multiple-model PHD, which considers different measurements models, rather than different mobility models, generally considered in the literature. 
        \item A new and theoretically sound method to update the vehicle state is provided, by deriving the closed form of the RFS-likelihood.
        \item By decomposing the overall source features into three mutually disjoint subsets, the fusion of each subset can be carried out independently. With this decomposition, we are able to flexibly design fusion weights for the different subsets and deal with non-overlapping sensor FoV. 
        \item Through a Gaussian mixture implementation with online source code, the efficacy of the proposed filter and fusion approach is demonstrated in a two-vehicle scenario \Ra{with 5G mmWave communication links}, where all propagation paths are exploited and vehicles cooperatively map the environment, which is shown to speed up the mapping process. 
    \end{itemize}
    The rest of the paper is organized as follows. In Section \ref{sec:RelatedWorks}, we discuss related works, and how our work compares to previous work. Section \ref{sec:Model} describes the considered vehicular networks with \Ra{5G mmWave  communication links} and a problem formulation. Section \ref{sec:LocalProcessingMMPHD} introduces the multiple-model PHD-SLAM at the vehicle. In Section \ref{sec:GlobalProcessingLKA}, asynchronous map fusion is presented. Numerical results and discussions are reported in Section \ref{sec:NumericalResults}. Finally, Section \ref{sec:Conclusions} concludes the paper.

\subsubsection*{Notation}
Throughout this paper, we will use the following basic notations. Scalars are denoted by italic, e.g., $x$.
Vectors are indicated by the bold lower-case letters, e.g., $\mathbf{x}$, and matrices are denoted by the bold upper-case letters, e.g., $\mathbf{X}$. Transpose of both vector and matrix is represented by superscript $\text{T}$, e.g., $\mathbf{x}^{\mathrm{T}}$ and $\mathbf{X}^{\mathrm{T}}$. Random sets are denoted by calligraphic, e.g., $\mathcal{X}$.
We denote probability density functions (pdfs) and probability mass functions (pmfs) by $f(\cdot)$ and $p(\cdot)$, respectively. We will use the following indexing: vehicle $n$, time step $k$, particle $i$, source type $m$, Gaussian mixture component $j$.

\section{Related Works} \label{sec:RelatedWorks}
In this section, we introduce the previous works for handling the aforementioned challenges in SLAM, multiple-model object tracking, and map fusion, considering methods based on geometry, methods based on message passing, and methods based on RFS theory.
%Previous solutions to SLAM can be coarsely categorized in to three classes: methods based on geometry \cite{yassin2018mosaic,aladsani2019leveraging,palacios2017jade,palacios2018communication}, methods based on message passing \cite{RicoTWC2018,HenkGlobecom2018,HyowonAsilomar2018}, and methods based on random finite set (RFS) theory \cite{mahler_AES_2003_PHD,Mullane2011,Leung2013,BATTISTELLI20172457,Deusch2015,DeuschRD:2015,FatemiGSRH:2016_PMBradarmapping}.

In the geometry-based SLAM methods, \cite{yassin2018mosaic} formulates the SLAM problem  using the geometric relation between observations, and a non-Bayesian estimator for the user location and extended Kalman filter for mapping are introduced.
MmWave imaging for one single reflected path is utilized in \cite{aladsani2019leveraging}.
Neither \cite{yassin2018mosaic} nor \cite{aladsani2019leveraging} considered the unknown number of objects or the data association uncertainty; in our paper we handle both an unknown number of objects and unknown data association.
The authors in \cite{palacios2017jade,palacios2018communication} develop SLAM methods that are applicable when the BS location is unknown. Their solutions require four anchors to localize a node regardless of whether the four anchors are physical BSs or VAs.
Therefore, at least one physical anchor is required with its corresponding VA mirrored through reflective surfaces for SLAM and obstacle detection.
%\Hyowon{Editor gave those corrections. Actually, they do not consider the SLAM challenges related to the unknown number of objects and uncertainty of data association. I am not sure that it is okay to mention them.}
% \Karl{How do we relate to this?}
% SLAM has been treated in a variety of applications including robotics \cite{thrun2005probabilistic}, location-awareness communication \cite{DiTaranto2014LAC}, map reconstruction \cite{engel2014lsd}, and radio-based positioning \cite{mendrzik2018joint,HenkGlobecom2018}.
% In \cite{GuiGueDar:J15,lemic2016localization,gunathillake2017topology,li2018massive}, 5G positioning and mapping in the indoor environment has been developed.
% However, this paper focuses on the 5G positioning and mapping for vehicular networks.

In the second category, a message passing-based estimator for position and orientation of the vehicle, as well as {mapping objects} is introduced in \cite{RicoTWC2018}.
In \cite{HenkGlobecom2018}, the clock bias of the vehicle is considered as an additional unknown, and scheduling method for effective message passing is introduced.
In \cite{mendrzik2018joint,HenkGlobecom2018}, only reflecting surfaces are regarded as objects generating multipath signals and small objects are ignored.
The authors in \cite{HyowonAsilomar2018} consider scatterers as well as reflection surfaces.
However, these message passing-based SLAM filters \cite{mendrzik2018joint,HenkGlobecom2018,HyowonAsilomar2018} do not include the data association uncertainty as part of the message passing problem, which we do in this paper.
For effective data association in message passing-based SLAM, the joint probability data association scheme is dealt with in \cite{meyer2018message}. 

The third approach involves RFS theory, which is a powerful tool for probabilistic modelling of a set of objects with uncertainties on both cardinality and object states. RFSs have been used for SLAM problems, see
\cite{mahler_AES_2003_PHD,Mullane2011,Leung2013,BATTISTELLI20172457,Deusch2015,DeuschRD:2015,FatemiGSRH:2016_PMBradarmapping}; these approaches mainly differ in terms of their representation of the object RFS and the required approximations. 
% \Hyowon{This part have been corrected for describing differentiation between the novelty offered by our paper and the literature. These corrected sentences are actions for Editor's comment ``However, I also agree that in your revision you should provide a better differentiation between the novelty offered by your paper and the literature, e.g., focused more on the different techniques used in your paper with respect to the literature, rather than on the 5G setting per se.'' and Reviewer 1's comment ``This paper introduces an integrated approach that consists of multiple components for positioning. However, the contribution and novelty are not very clear. It seems that each component is a variation or adaptation of existing methods, and the paper integrates a long list of them.''.}
When using an RB SLAM density with an RFS based map, the position particle weights must be updated. In previous work, the RFS-likelihood for vehicle position update was approximated by using a dummy map, where the dummy map was either an empty map or a map with a single feature \cite{Mullane2011,Leung2013}. In comparison, in this paper we use the theoretically exact RFS-likelihood.
Unlike this work, where a PHD representation of the map RFS density is used, the authors of \cite{Deusch2015,DeuschRD:2015,FatemiGSRH:2016_PMBradarmapping} represent the RFS density different types of densities (e.g., multi-Bernoulli (MB) and Poisson MB). However, those RFS densities require explicit data association, which is less computationally efficient.
Further, in \cite{FatemiGSRH:2016_PMBradarmapping} the authors only consider the mapping problem, which is a simpler problem compared to SLAM.

In object tracking, multiple models are commonly used to handle maneuvering targets that switch between different types of motion, e.g., going straight forward or turning, see, e.g., \cite{BlomBS1988}, however it is also possible to use multiple models to handle different types of measurements, see, e.g., \cite{GranstromL:2013,GranstromRMS:2014}. The objects can transition from one type of motion to another type, and this is commonly modelled using a jump Markov system, which can be handled, e.g., using the interactive multiple model (IMM) estimator \cite{BlomBS1988}. However, in the considered 5G SLAM application, the objects do not transition from one type to another, and subsequently jump Markov system modelling is not necessary.
There are multiple ways in which one can integrate multiple models into a PHD filter; an overview of different approaches to multiple model PHD filters was given in \cite{Mahler:2012}. In this paper we adopt the proper jump Markov chain model in \cite{Pasha2009} which is based on an augmented object state consisting of the object state and the object type.

Fusion of the different map PHDs from different vehicles defined in the different and limited FoVs brings considerable challenges.
Generally speaking, two frameworks can be employed in this situation: (i) centralized methods, where each vehicle directly sends the raw measurements to the fusion center to perform SLAM; (ii) decentralized methods, where each vehicle process the measurements and then share their posteriors with  each other (or a fusion center) to perform density fusion. The centralized method is computationally intensive for the BS and treats the vehicles as decentralized sensors. To spread out the complexity over the network, the focus has been on decentralized methods. 
The most prevalent methods for multi-object density fusion consider generalized covariance intersection (GCI) \cite{Li2018MSMOFOV,li2019distributed}, which amounts to computing the intersection of information among densities. GCI cannot be directly applied in 5G cooperative SLAM because of the multi-object densities are defined for different FoVs. This difficulty is overcome in \cite{BATTISTELLI20172457}, where the PHD of each vehicle is initialized as a non-zero constant throughout the whole area of interest. Though it works well in fusing maps with different FoVs, 
%the non-zeros PHD of everywhere imposes a huge amount of computational load. 
it becomes troublesome when applied to large-scale scenarios since the storage and the propagation of the resulting PHD, which is non-zero everywhere, require both large memory and computational resources. In addition, GCI in \cite{Li2018MSMOFOV,li2019distributed} extracts minimum information in fusing the maps. Thus we adopt AA which takes the union of involved densities and leads to minimum information loss  \cite{li2017generalized,li2018local,Gao2019multiobject}. However, there is a challenge of selecting the fusion weight in our scenario.

None of the above methods have been applied to the problem of 5G mmWave cooperative positioning and mapping. 

\section{Model}\label{sec:Model}
In this section, we describe a vehicle, environment, and measurement models for the considered \Ra{propagation environment with 5G mmWave communication links.} %state and vehicle dynamics. } With considering the 5G properties we introduce a 5G propagation environment, for which a observation is modeled.}

\subsection{Vehicle Model} \label{sec:VehicleEnvironmentModel}
We consider a set of $N$ vehicles, traversing a common environment, in communication with a common BS. The BS has a known and fixed location $\mathbf{x}_{\text{BS}}\in\mathbb{R}^{3}$.  Each vehicle $n$ has a dynamic state $\mathbf{s}^{(n)}_{k}$ at time $k$. Time is discrete with sampling interval $\Delta$. The state comprises the three-dimensional position $\mathbf{v}^{(n)}_{k}=[x^{(n)}_{k}, y^{(n)}_{k} z^{(n)}_{k}]^\textrm{T}$, heading ${\alpha}^{(n)}_{k}$, {translation speed} $\zeta^{(n)}_{k}$, {turn-rate} $\rho^{(n)}_{k}$, and clock bias ${B}^{(n)}_{k}$. Vehicle $n$ has a known dynamic model {with the transition density} ${f}(\mathbf{s}^{(n)}_{k}|\mathbf{s}^{(n)}_{k-1})$. The vehicle dynamics follow a velocity motion model 
\begin{align}\label{eq:processModel}
\mathbf{s}^{(n)}_{k} & = \bm{\upsilon}(\mathbf{s}^{(n)}_{k-1}) + \mathbf{q}_{k}, 
%& \mathbf{s}^{(n)}_{k}=\mathbf{s}^{(n)}_{k-1}+ \\
%& \left[\begin{array}{c}
%\frac{\zeta^{(n)}_{k-1}}{\rho^{(n)}_{k-1}}\left(\sin(\alpha^{(n)}_{k-1}+\rho^{(n)}_{k-1}\Delta)-\sin\alpha^{(n)}_{k-1}\right)\\
%\frac{\zeta^{(n)}_{k-1}}{\rho^{(n)}_{k-1}}\left(-\cos(\alpha^{(n)}_{k-1}+\rho^{(n)}_{k-1}\Delta)+\cos\alpha^{(n)}_{k-1}\right)\\
%0\\
% \rho^{(n)}_{k-1}\Delta\\
%\mathbf{0}_3
%\end{array}\right]+\mathbf{q}_{k}, \nonumber
\end{align}
where $\bm{\upsilon}(\cdot)$ is a known transition function (see \cite[Chapter 5]{thrun2005probabilistic}, \cite{RongLiJ:2003} and Section \ref{sec:NumericalResults}) and 
%$\mathbf{0}_\psi$ denotes a column vector of $\psi$ zeros, $\Delta$ is the \textcolor{magenta}{sampling time} and  
$\mathbf{q}_{k}$ denotes the process noise, modeled as zero-mean Gaussian with known covariance  $\mathbf{Q}$.

%\begin{figure}
%\centering\includegraphics[width=.4\columnwidth]{Figures/Scenario_V2.pdf}
%\caption{{Scenario with the environment (BS, VA, and SP) and two vehicles (heading $\alpha$ shown with an arrow and clock bias $B$). The domes represent the SP field of view of the vehicles, here modeled as a half-sphere with radius $r_\textrm{FoV}$.}}
%\label{fig:scenario123}
%\end{figure}

\subsection{Environment Model}

 The environment is characterized by scattering points (SPs) and reflecting surfaces. % (see Fig.~\ref{fig:scenario123}). 
 A scattering point has an unknown three-dimensional location $\mathbf{x}_{\text{SP}}$, while a reflecting surface can be parameterized by a fixed virtual anchor
(VA) location $\mathbf{x}_{\mathrm{VA}}$, obtained by mirroring\footnote{Mathematically, the reflecting surface can be described by a point
$\mathbf{f}$ and a normal vector $\mathbf{u}$. With each
reflecting surface we can associate a virtual anchor location $\mathbf{x}_{\mathrm{VA}}=\mathbf{P}\mathbf{x}_{\text{BS}}+\mathbf{t}$,
where $\mathbf{P}=\mathbf{I}_{3}-2\mathbf{u}\mathbf{u}^{\mathrm{T}}$
is a Householder matrix and $\mathbf{t}=2\mathbf{f}^{\mathrm{T}}\mathbf{u}\mathbf{u}$
is a translation vector.} the BS with respect to the surface. The details of geometric relation to the propagation environment are described in Appendix \ref{sec:Geometry}.

\subsection{Observation Model}\label{sec:ObservationModel}
A common model of a \Ra{5G mmWave} received signal from the BS to vehicle $n$ at time $k$ is {\cite{Heath2016}}
\begin{align}\label{eq:ObservationModel1}
\mathbf{y}^{(n)}_k(t)= 
(\mathbf{W}^{(n)}_k)^{\text{H}}\sum_{l=0}^{L^{(n)}_{k,l}}h^{(n)}_{k,l}\mathbf{a}_{\text{R}}(\bm{\theta}^{(n)}_{k,l})\mathbf{a}_{\text{T}}^{\text{H}}(\bm{\phi}^{(n)}_{k,l})\mathbf{b}_k(t-\tau^{(n)}_{k,l})+\mathbf{n}^{(n)}_{k}(t), 
\end{align}
where $\mathbf{b}_k(t)$ is a transmitted signal (possibly precoded) to all the users,
$\mathbf{W}^{(n)}_k$ is a combining matrix, $h^{(n)}_{k,l}$ is a complex path gain,
$\bm{\theta}^{(n)}_{k,l}$ is the AOA (also denoted as direction
of arrival - DOA) in azimuth and elevation, $\bm{\phi}^{(n)}_{k,l}$ is the AOD (also denoted as direction of departure -
DOD) in azimuth and elevation, $\tau^{(n)}_{k,l}$ is the time of arrival
(TOA), and $\mathbf{n}^{(n)}_k(t)$ is (possibly colored) noise. The vectors $\mathbf{a}_{\text{T}}^{\text{H}}(\bm{\phi})$ and $\mathbf{a}_{\text{R}}^{\text{H}}(\bm{\theta})$ are the steering vectors of the transmit and receive array, respectively.  The AOA and
TOA are measured in the frame of reference of the receiver, while
the AOD is measured in the frame of reference of the transmitter.
The path index $l=0$ is the line-of-sight (LOS) path, while the $L^{(n)}_{k,l}$
remaining paths are non-LOS (NLOS) paths. The AOA, TOA, and AOD of
each path has a geometric meaning, which depends on the location of
the transmitter and receiver, as well as the points of incidence of
the NLOS paths in the environment (see further). 
We further assume a channel estimation routine is present
at the receiver, which provides, at time $k$, a set $\mathcal{Z}^{(n)}_k$ of measurements with elements
\begin{align}
&\mathbf{z}^{(n)}_{k,l}=\mathbf{h}(\mathbf{x},\mathbf{s}^{(n)}_k,m)+\mathbf{w}^{(n)}_{k,l},
\label{eq:ObservationModel2}
\end{align}
where
\begin{align}\label{eq:ObservationModel2b}
\mathbf{h}(\mathbf{x},\mathbf{s}^{(n)}_k,m) = [\tau^{(n)}_{k,l}\,,(\bm{\theta}^{(n)}_{k,l})^{\text{T}},(\bm{\phi}^{(n)}_{k,l})^{\text{T}}]^{\text{T}}
\end{align}
 and $\mathbf{w}_{k,l}\sim\mathcal{N}(\mathbf{0},\bm{\Sigma}^{(n)}_{k,l})$ for a certain number of paths $l=0,\ldots,L^{(n)}_{k}$. Here, $m$ denotes the source type and $\mathbf{x}$ the source location. We distinguish between three different sources: the BS, a VA, a SP and correspondingly have  $m\in\{\mathrm{BS},\mathrm{SP},\mathrm{VA}\}$. Both the source type and source location are unknown. We define $\mathcal X$ as a random set of sources with entries $(\mathbf{x},m)$ with density $f(\mathcal{X})$. The functional form of $\mathbf{h}(\mathbf{x},\mathbf{s}^{(n)}_k,m)$ and of the corresponding likelihood function $f(\mathbf{z}^{(n)}_l | \mathbf{x},\mathbf{s}^{(n)}_k, m)$ is described in detail in Appendix \ref{sec:Geometry}. 
 
 Finally, not all sources give rise to measurements and some measurements don't correspond to any fixed source. This is described as follows: 
 \begin{itemize}
     \item \emph{Missed detections:} A vehicle may only be able to detect a source if it is within the field of view. Hence, we introduce $p_{\text{D},k}(\mathbf{x},\mathbf{s}^{(n)}_k,m)\in [0,1]$ as the probability that a source of type $m$ with location $\mathbf{x}$  can give rise to a measurements when the vehicle is in state $\mathbf{s}^{(n)}_k$.
     \item \emph{False alarms:} Some measurements in $\mathcal{Z}^{(n)}_k$ may  correspond to clutter (e.g., due to noise peaks that are detected as paths during channel estimation). We model this through the clutter intensity   $c(\mathbf{z})$, which assumes that clutter is generated according to a Poisson point process.  
     \item \emph{Transient sources:} Measurements may also correspond to transient physical objects in the environment (e.g., a vehicle that moves). The corresponding measurements can be seen as a landmark that is visible only for a short time (a few seconds) and will be treated as a transient SP, meaning that it will appear and then disappear from the map. 
 \end{itemize}
 We assume that $\mathbf{h}(\mathbf{x},\mathbf{s}^{(n)}_k,m)$, $\bm{\Sigma}^{(n)}_{k,l}$, $p_{\text{D},k}(\mathbf{x},\mathbf{s}^{(n)}_k,m)$ and $c(\mathbf{z})$ are known to vehicle $n$.

\subsection{Problem Formulation}
Given a certain prior $f(\mathbf{s}^{(n)}_{0})$, our goal
is to track the state of the vehicles' states and build a common map of the environment  (VAs and SPs). To solve this problem, we first detail the SLAM algorithm running locally on each vehicle and then go on to detail the map fusion at the BS. 
% \textcolor{blue}{(it is a bit too brief. For example, what is the map state)}

\section{Local processing: multiple-model PHD-SLAM}\label{sec:LocalProcessingMMPHD}
In this section, we describe a local multiple-model PHD filter at each vehicle.
%The multiple-model PHD filter is proposed for the local process.
We will consider a single vehicle and thus drop the vehicle index $n$.

\subsection{Approach}
The map state will be modeled as a multi-object Poisson process (MPP), which is fully characterized by its PHD (first-order statistical moment), hence the conditional map PHD is propagated rather than its density. Further, in order to distinguish the type of each source, the discrete state $m$ is also included in conditional map PHD. We rely on the standardized RB approach, whereby the vehicle state trajectory is represented by particles, and PHDs conditioned on each particle are maintained. 
Hence, the data structure at the end of time $k-1$ {consists of} (i) a list of $I$ particles $\mathbf{s}^{i}_{0:k-1}$ with {particle} weights $\omega^{i}_{k-1|k-1}\ge 0$, $\sum_i \omega^{i}_{k-1}=1$; (ii) for each particle, the PHD $D_{k-1|k-1}(\mathbf{x},m|\mathbf{s}^{i}_{0:k-1})$, $m\in\{\mathrm{BS},\mathrm{SP},\mathrm{VA}\}$.
We initialize $D_{0|0}(\mathbf{x},\text{VA}|\mathbf{s}^{i}_{0})=D_{0|0}(\mathbf{x},\text{SP}|\mathbf{s}^{i}_{0})=0$ and  $D_{0|0}(\mathbf{x},\text{BS}|\mathbf{s}^{i}_{0}) = \delta(\mathbf{x}-\mathbf{x}_{\text{BS}})$.
As a shorthand, we will denote $D_{k-1|k-1}(\mathbf{x},m|\mathbf{s}^{i}_{0:k-1})$ as $D^{i}_{k-1|k-1}(\mathbf{x},m)$, $p_{\text{D},k}(\mathbf{x},\mathbf{s}^{(n,i)}_k,m)$ as $p^i_{\text{D},k}(\mathbf{x},m)$, and $\mathbf{h}(\mathbf{x},\mathbf{s}^{(n,i)}_k,m)$ as $\mathbf{h}^i(\mathbf{x},m)$.

We are now ready to describe the recursive formulation of the PHD filter (Section \ref{sec:PHDtheory}), followed by a practical Gaussian mixture  implementation (Section \ref{sec:PHDGM}). 

\subsection{Basics on PHDs}
An RFS $\mathcal{X}$ is characterized by its set density $f(\mathcal{X})$, which in turn depends on the cardinality distribution and the cardinality-conditioned joint distributions  \cite{williams2015marginal}
\begin{align}
f(\{\mathbf{x}_1,\mathbf{x}_2,\ldots, \mathbf{x}_n\})=p(n)\sum_{\pi}f_n(\mathbf{x}_{\pi(1)},\ldots, \mathbf{x}_{\pi(n)}),  
\end{align}
where $p(n)$ is the cardinality distribution evaluated in $n$,  the sum goes over all permutations $\pi$ of the set $\{1,\ldots, n\}$, and $f_n()$ is standard vector density of $n$ elements. The set integral is defined as
\begin{align}\label{eq:setIntegral}
    \int g(\mathcal{X}) \delta \mathcal{X} =  g(\emptyset) + \sum_{{n=1}}^{{\infty}}\frac{1}{n!}\int g(\{{\mathbf{x}_1,\mathbf{x}_2,\ldots, \mathbf{x}_n}\})\text{d}\mathbf{x}_1\ldots \text{d}\mathbf{x}_n.
\end{align}
If $\delta_\mathcal{X}(\mathbf{x})=\sum_{\mathbf{w}\in \mathcal{X}}\delta_\mathbf{w}(\mathbf{x})$, where $\delta(\cdot)$ indicates the delta Dirac function, the PHD associated with $\mathcal{X}$ is the function \cite{mahler_AES_2003_PHD}
\begin{align}
     D({\mathbf{x}})  = \int f(\mathbf{x} \cup \mathcal{W})\delta \mathcal{W} = \int \delta_\mathcal{X}(\mathbf{x})f(\mathcal{X}) \delta  \mathcal{X},
\end{align}
which has as property that for any region $S$ in the underlying state space, $\int_S D(\mathbf{x}) \text{d}\mathbf{x}$ is the expected number of elements in $S$. Note that $D(\mathbf{x})\ge {0}$ is generally not normalized, and generally does not provide a unique representation of an RFS density (multiple RFS densities may have the same PHD). One exception is the Poisson Point Process (PPP) RFS, which has a single parameter, called the PPP intensity, which is equal to the PPP PHD. In this case the RFS density is defined as follows, see, e.g., \cite{Mullane2011},
\begin{align}
    f(\mathcal{X}) = \frac{\prod_{\mathbf{x} \in \mathcal{X}}D(\mathbf{x})}{\exp(\int D(\mathbf{x})\text{d}\mathbf{x})}. \label{eq:PHDtoPDF}
\end{align}

A common representation of a PHD is through a Gaussian mixture (GM)
\begin{align}
    D(\mathbf{x}) \approx \sum_{j=1}^{J} \gamma^j \mathcal{N}(\mathbf{x};\bm{\mu}^j,\bm{\Sigma}^j),
\end{align}
where $\sum_{j=1}^{J} \gamma^j$ represents the expected number of elements, with locations $\bm{\mu}^j$. The GM representation allows closed form computation of the PHD mapping filter under certain conditions. 
%The vehicle will rely on {a single PHD-filter with a state that consists of both a continuous source position vector $\mathbf{x}$ and a discrete source type scalar $m$,} and a particle representation of the vehicle state trajectories. 

%(At each vehicle, the joint posterior is represented by a set of particles, where each particle includes the trajectory of the vehicle and the map density conditioned on the vehicle trajectory. In this paper, 

%\GB{In the current model, the set of sources \[ \mathcal X = \{ (x_1,m_1) , \ldots, (x_n,m_n) \}\] consisting of BS, VAs and SPs is modelled as a Poisson point process with intensity $D(x,m)$. This is perhaps the simplest way to address the considered scenario and, hence, I agree that it makes sense to consider this model as a first attempt. On the other hand, this model does not account for the fact that the set $\mathcal X$ always contains exactly one element with label BS and position $x_{\rm BS}$. An alternative would be to model $\mathcal X$ as the union of the BS $\{ (x_{\rm BS},{\rm BS}) \}$ and a Poisson point process, say $\bar{\mathcal X}$, consisting only of VAs and SPs. However, this would require to work out the equation of the multi-target Bayes filter for this case.}

\subsection{General Formulation} \label{sec:PHDtheory}
The filter comprises two steps: the prediction step, which accounts for the motion model \eqref{eq:processModel}, and the update step, which accounts for the measurement set $\mathcal{Z}_k$. 
\subsubsection{Prediction}
  The PHD prediction is \cite{Vo2006}
\begin{align}\label{eq:MapPredictionModel}
D^{i}_{k|k-1}(\mathbf{x},m)=D^{i}_{k-1|k-1}(\mathbf{x},m)+b^{i}(\mathbf{x},m),    
\end{align}
where $b^{i}(\mathbf{x},m)$ is a birth process, indicating where and with which intensities we expect sources of type $m$ to appear. Note that $b^{i}(\mathbf{x},\text{BS})=0$ since the BS location is already known.
For the vehicle state prediction, we use the process model {\eqref{eq:processModel}} to generate predicted trajectories, $\mathbf{s}^{i}_{0:k}$, where $\mathbf{s}^{i}_{k} \sim f(\mathbf{s}_{k}|\mathbf{s}^{i}_{k-1})$, with  $\omega^{i}_{k|k-1} = \omega^{i}_{k-1|k-1}$.

\subsubsection{Measurement Update}
Given the measurement set $\mathcal{Z}_{k}$ at time $k$, {we update the 3 PHDs for each particle as follows: for the BS PHD, $D^i_{k|k}(\mathbf{x},\text{BS}) = D^i_{k|k-1}(\mathbf{x},\text{BS})$, which for the VA and SP PHDs \cite{Vo2006}},
\begin{align}\label{eq:MapUpdateModel}
D^i_{k|k}(\mathbf{x},m)  =D^i_{k|k-1}(\mathbf{x},m)(1-p^i_{\text{D}}(\mathbf{x},m))+\sum_{\mathbf{z}\in\mathcal{Z}_{k}}\frac{\nu^i(\mathbf{z},\mathbf{x},m)}{c(\mathbf{z}) + \sum_{m'}\int \nu^i(\mathbf{z},\mathbf{x}',m')\text{d}\mathbf{x}'}, 
\end{align}
where we recall that  $c(\mathbf{z})$ is the
clutter intensity, $p^i_{\text{D}}(\mathbf{x},m)$ is shorthand for the detection probability of a source of type $m$ at location $\mathbf{x}$ (given the current vehicle state $\mathbf{s}^i_k$) and
\begin{align}
 \nu^i(\mathbf{z},\mathbf{x},m)  = p^i_{\text{D}}(\mathbf{x},m)f(\mathbf{z}|\mathbf{x},\mathbf{s}^i_k,m)D^i_{k|k-1}(\mathbf{x},m). \label{eq:nu-function}
\end{align}
The first term in \eqref{eq:MapUpdateModel} corresponds to the update when no measurement comes from the source at location $\mathbf{x}$ (as it is out of the field of view), while the second term corresponds to the update when there is a measurement. In the latter case, the measurement can come from clutter, which is accounted for in the denominator. 

In parallel, using the same measurement set $\mathcal{Z}_{k}$, we update the vehicle state distribution, but updating the weights 
%\GB{In view of Remark 1, perhaps it makes sense to let the clutter intensity depend also on the vehicle state $s^i_k$, i.e. to have $c^i (z) = c(z|s^i_k)$, so as to take into account the information on the vehicle state provided by the first type of clutter (i.e. landmarks that are visible only for a short time). In fact, I remember discussing with Henk that his information could be useful to improve localization accuracy (for instance to compute the weights in equation (8)). I think that this depends also on the birth process that we use.}
%\subsubsection{Vehicle state}
%Given the same measurements, the weight of particle $i$ is updated as 
\begin{align}
     \omega^{i}_{k|k}   =\omega^{i}_{k|k-1} f(\mathcal{Z}_k|\mathbf{s}^i_{0:k},\mathcal{Z}_{1:k-1}) = \omega^{i}_{k|k-1} \int  f(\mathcal{Z}_k|\mathbf{s}^i_{0:k},\mathcal{X},\mathcal{Z}_{1:k-1})f(\mathcal{X}|\mathbf{s}^i_{0:k},\mathcal{Z}_{1:k-1}) \delta \mathcal{X}, \label{eq:vehStateUpdate}
\end{align}
where $\delta \mathcal{X}$ refers to a set integral. To avoid numerical problems, rather than working with the particle weights $\omega_{k|k}^{i}$, we work with the log-weights $\ell_{k|k}^{i} = \log\omega_{k|k}^{i}$. The log-weight update is
$    \ell_{k|k}^{i} =  \ell_{k|k-1}^{i} + \log f(\mathcal{Z}_k|\mathbf{s}^i_{0:k},\mathcal{Z}_{1:k-1})$.
    %\\= & \ell_{k|k-1}^{i} + \sum_{\mathbf{z}\in\mathcal{Z}} \log\left( c(\mathbf{z}) + \sum_{m'}\int \nu^i(\mathbf{z},\mathbf{x}',m')\text{d}\mathbf{x}' \right)

In previous work on PHD-SLAM \cite{Mullane2011,Leung2013}, the integral in the weight update \eqref{eq:vehStateUpdate}, 
\begin{align}
    \int  f(\mathcal{Z}_k|\mathbf{s}^i_{0:k},\mathcal{X},\mathcal{Z}_{1:k-1})f(\mathcal{X}|\mathbf{s}^i_{0:k},\mathcal{Z}_{1:k-1}) \delta \mathcal{X}
\end{align}
was approximated using a ``dummy'' map $\hat{\mathcal{X}}$; in \cite[Sec. 4.E]{Mullane2011} it is proposed to use either an empty map or a map with a single feature, in \cite[Sec. 3.C]{Leung2013} a map with multiple features is used.
In this paper, we use the exact expression for the integral in \eqref{eq:vehStateUpdate}. With a PPP prior $f(\mathcal{X}|\mathbf{s}^i_{0:k},\mathcal{Z}_{1:k-1})$ and a point object measurement model, $f(\mathcal{Z}_k|\mathbf{s}^i_{0:k},\mathcal{X},\mathcal{Z}_{1:k-1})$,  the solution to the integral is
\begin{align} \label{eq:likelihood}
    f (\mathcal{Z}_k|\mathbf{s}^i_{0:k},\mathcal{Z}_{1:k-1})  =
     \prod_{\mathbf{z}\in\mathcal{Z}_k} \left( c(\mathbf{z}) + \sum_{m'}\int \nu^i(\mathbf{z},\mathbf{x}',m')\text{d}\mathbf{x}' \right).
\end{align} 
This result follows as a special case of the more general PMBM update, see details in \cite[Sec. 3.B.2]{Garcia-Fernandez2018}, as derived in Appendix \ref{sec:weightProof}. Note that \eqref{eq:likelihood} is easily evaluated during the map update step.

\subsection{Gaussian Mixture Implementation} \label{sec:PHDGM}
While the expression above provide a solution to the SLAM problem, considering multiple source types and limited field of view, a practical implementation requires several choices and approximations to be made. In this section, we provide a GM implementation, inspired by \cite{Mullane2011}. The proposed implementation has a complexity cost that scales as $\mathcal{O}(I \times N_{\text{model}} \times J \times  |\mathcal{Z}|)$ per vehicle and per time step. Here, $I$ denotes the number of particles, $N_{\text{model}}$ is the number of models, $J$ is the number of Gaussian mixture components per model, and $|\mathcal{Z}|$ is the number of measurements per time step. Note that for single-model SLAM $N_{\text{model}}=1$, for mapping only $I=1$, and for localization only, $N_{\text{model}} \times J=1$.

 Using a set of $I$ particles, the multiple-model PHD-SLAM density at time $k-1$ is expressed as
 \begin{align}
  \{\mathbf{s}^{i}_{k-1|k-1},\omega^{i}_{k-1|k-1}, D^{i}_{k-1|k-1}(\mathbf{x},m)\}_{i=1}^{I},
 \end{align}
 where $D^{i}_{k-1|k-1}(\mathbf{x},m)$ will be described by a GM. In this section, we will detail the implementation of map prediction \eqref{eq:MapPredictionModel}, map update \eqref{eq:MapUpdateModel}, and vehicle state update \eqref{eq:vehStateUpdate}.
 %\HK{The BS map is already known, and there are no birth in the BS map. Should that facts be mentioned?} --> fixed in previous section

\subsubsection{Map Prediction (\ref{eq:MapPredictionModel})}
The map PHD $D^{i}_{k-1|k-1} (\mathbf{x},m)$ at the end of time step $k-1$ is assumed to be of a GM
\begin{align}\label{eq:CorrectionMapGM}
    D^{i}_{k-1|k-1}(\mathbf{x},m) = \sum_{j=1}^{J_{k-1|k-1}^{i}(m)}\gamma^{i,j}_{k-1|k-1}(m)\mathcal{N}(\mathbf{x};\mathbf{x}^{i,j}_{k-1|k-1}(m),\mathbf{P}^{i,j}_{k-1|k-1}(m)),
\end{align}
where $J_{k-1|k-1}^{i}(m)$ is the number of Gaussians in the map PHD for the source type $m$, and $\gamma^{i,j}_{k-1|k-1}(m) \ge 0$, $\mathbf{x}^{i,j}_{k-1|k-1}(m)$, and $\mathbf{P}^{i,j}_{k-1|k-1}(m)$ are respectively the weight, mean, and covariance of $j$-th Gaussian. Note that 
$\sum_{j}\gamma^{i,j}_{k-1|k-1}(m)$ is not necessary to be equal to 1.
Similarly, the birth process PHD $b^{i}_{k}(\mathbf{x},m)$, which is determined as the measurement arrives, is also represented as a GM
    \begin{align}\label{eq:BirthProcessGM}
        b^{i}_{k}(\mathbf{x},m) &
        = \sum_{j(\mathbf{z})=1}^{J_{b,k}^{i}(m)}\gamma^{i,j}_{b,k}(m)\mathcal{N}(\mathbf{x};\mathbf{x}^{i,j(\mathbf{z})}_{b,k}(m),\mathbf{P}^{i,j(\mathbf{z})}_{b,k}(m)),
    \end{align}
% where $J_{b,k}^{i}(m)$ is the number of Gaussians {in} the birth process PHD, $\mathbf{x}^{i,j}_{b,k}(m)$ and $\mathbf{P}^{i,j}_{b,k}(m)$ are respectively the mean and covariance of Gaussians which indicate the statistics of the birth location.
where $j(\mathbf{z})$ denote the measurement index corresponding to the measurement $\mathbf{z}$, and $J_{b,k}^{i}(m)$ is the number of Gaussians {in} the birth process PHD, which is equal to the number of elements in measurement $\mathcal{Z}_k$. Hence, $\mathbf{x}^{i,j(\mathbf{z})}_{b,k}(m)$ and $\mathbf{P}^{i,j(\mathbf{z})}_{b,k}(m)$ are respectively the mean and covariance of Gaussians which indicate the statistics of the birth location.
Hence, the prediction map PHD $D^{i}_{k|k-1}(\mathbf{x},m)$ in \eqref{eq:MapPredictionModel} is  given by the sum of $D^{i}_{k-1|k-1}(\mathbf{x},m)$ and birth process PHD $b^{i}_{k}(\mathbf{x},m)$, which is a new GM, denoted by 
\begin{align}\label{eq:PredictionGMPHD}
    D^{i}_{k|k-1}(\mathbf{x},m) = 
    \sum_{j=1}^{J_{k|k-1}^{i}(m)}\gamma^{i,j}_{k|k-1}(m)\mathcal{N}(\mathbf{x};\mathbf{x}^{i,j}_{k|k-1}(m),\mathbf{P}^{i,j}_{k|k-1}(m)),
\end{align}
where $J_{k|k-1}^{i}(m)=J_{k-1|k-1}^{i}(m)+J_{b,k}^{i}(m)$.

An important practical consideration is how to set the weights, means and covariances of the birth process. We have found that in order to have an implementation that is able to successfully incorporate new information, it was crucial to let these depend on the measurements at time $k$, so that $\nu^i(\mathbf{z},\mathbf{x},m)$ in \eqref{eq:nu-function} takes on significant values \cite{Ristic2012}. 
% The main idea is, for each measurement $\mathbf{z}$, to generate a source (a birth) in each map. The mean $\mathbf{x}^{i,j}_{k|k-1}(\mathbf{z},m)$ and covariance 
% $\mathbf{P}^{i,j}_{k|k-1}(\mathbf{z},m)$
% of these sources can be determined by the inverse sigma point of the cubature Kalman filter (CKF) \cite{HaykinCKF2009}, details of which are described in Appendix \ref{sec:GMofBirth}.
The main idea is, for each measurement $\mathbf{z}$, to generate a birth for each source type $m$. Using the inverse sigma point of the cubature Kalman filter (CKF) \cite{HaykinCKF2009}, the mean $\mathbf{x}^{i,j}_{b,k}(m)$ and covariance 
$\mathbf{P}^{i,j}_{b,k}(m)$
of these births can be determined with respect to the measurement $\mathbf{z}$ and source type $m$, details of which are described in Appendix \ref{sec:GMofBirth}.
The weight $\gamma^{i,j}_{k|k-1}(m)$
is set to a low constant value, depending on the application.
Complexity can be reduced by not generating sources with low likelihood (e.g., when the generated source location is out of the field of view so that $p^i_{\text{D}}(\mathbf{x}^{i,j}_{k|k-1}(m),m)$ close to zero).

\subsubsection{Map Update (\ref{eq:MapUpdateModel})}
In order to evaluate the update in closed form, we utilize two approximations: the first approximation involves the detection probability and the second approximating the Bayes update.
% We note that since the births are generated from the measurements, their detection probability should be 1 and they should not be updated with any measurement \cite{Ristic2012}.
We note that since the births are generated from the measurements, their detection probability should be 1 and
they should not be updated with their corresponding measurements (i.e., the likelihood for a birth and its corresponding measurement is set to 1) \cite{Ristic2012}. 
{For the existing targets, on the other hand, we consider an adaptive detection probability $p^{i,j}_{\text{D},k}(m)$. We may set this adaptive detection probability to the expected value (i.e., $p^{i,j}_{\text{D},k}(m) = \mathbb{E}\{ p^{i,j}_{\text{D},k}(\mathbf{x},m)\}$ where the expectation is over $\mathbf{x}$ with density $\mathcal{N}(\mathbf{x};\mathbf{x}^{i,j}_{k|k-1}(m),\mathbf{P}^{i,j}_{k|k-1}(m))$) or to a robust value to avoid weight decrease of objects that were previously detected (i.e., $p^{i,j}_{\text{D},k}(m)=\min_{\mathbf{x} \in \mathcal{S}}
~p_{\text{D},k}^{i,j}(\mathbf{x},m)$, where $\mathcal{S}$ could be  the highest density region of  $\mathcal{N}(\mathbf{x};\mathbf{x}_{k|k-1}^{i,j}(m),\mathbf{P}_{k|k-1}^{i,j}(m))$ containing a large fraction (e.g., 95\%)
%\footnote{In case $p^{i,j}_{\text{D},k}(\mathbf{x},m)=p_{\text{D}}$ \textcolor{magenta}{when $\Vert \mathbf{x}- \mathbf{v}_k^{i}\Vert+\epsilon < r_{\text{FoV}}$} and $0$ otherwise, where $r_{\text{FoV}}$ denotes the radius corresponding to the FoV, $\mathbf{v}_k^{i}$ is the position of the $i$-th particle, and \textcolor{magenta}{$\epsilon$ indicates object uncertainty which can be represented by the object covariance.} Denoting by $\mathbf{e}$ the unit-length vector from $\mathbf{v}_k^{i}$ to $\mathbf{x}^{i,j}_{k|k-1}(m)$, then $d=\mathbf{e}^\text{T}(\mathbf{x}-\mathbf{v}_k^{i})$ with $d  \sim \mathcal{N}(d;\Vert \mathbf{x}^{i,j}_{k|k-1}(m)-\mathbf{v}_k^{i}\Vert,\mathbf{e}^\text{T}\mathbf{P}^{i,j}_{k|k-1}(m)\mathbf{e})$ from which it immediately follows that, with  $\sigma^2=\mathbf{e}^\text{T}\mathbf{P}^{i,j}_{k|k-1}(m)\mathbf{e}$ and $\bar{d}=\Vert \mathbf{x}^{i,j}_{k|k-1}(m)-\mathbf{v}_k^{i}\Vert$. Then, we can further approximate $p^{i,j}_{D,k}(m)$ as  \textcolor{magenta}{$p^{i,j}_{\text{D},k}(m) \approx p_{\text{D}} u(r_\text{FoV}-(\bar{d}+3\sigma))$, where $u(\cdot)$ is denoted by the unit step function.}} 
of the mass). %While this will lead to more false alarms and fewer missed detections, it turns out to give better results. %The object by the false alarms is excluded in the map fusion, will be described in Sec. \ref{sec:UplinkTransmission}.
}
{PHD filters are known for being sensitive to both missed detections and false alarms, due to the approximation of the multi-object density as a Poisson RFS. The Poisson cardinality has high variance, so a missed detection leads to a drastic decrease in the landmark weight (except when the detection probability is very low), while clutter often leads to false landmarks. Hence, if we don't want to lose the sources due to missed detections, we must set the detection probability to low values, at a cost of  a higher sensitivity to clutter (false landmarks). }

The second approximation is related to the Bayes update, and %\textcolor{magenta}{update of the sequential Bayesian filtering \cite{Sarkka2013}} to 
allows a closed-form evaluation of \eqref{eq:nu-function}
\begin{align}\label{eq:nu-sum}
  \nu^i(\mathbf{z},\mathbf{x},m)  &= \sum_{j=1}^{J_{k|k-1}^{i}(m)} \nu^{i,j}(\mathbf{z},\mathbf{x},m).
\end{align}
% where for a birth $j$, \cite{Ristic2012}
%   \begin{align}\label{eq:nu-birth}
%       \nu^{i,j}(\mathbf{z},\mathbf{x},m)= \gamma^{i,j}_{k|k-1}(m)\mathcal{N}(\mathbf{x};\mathbf{x}^{i,j}_{k|k-1}(m),\mathbf{P}^{i,j}_{k|k-1}(m)), 
%   \end{align}
%   while for $j$ not corresponding to a birth
%   \begin{align}
%   \nu^{i,j}(\mathbf{z},\mathbf{x},m) &= \gamma^{i,j}_{k|k-1}(m)p^{i,j}_{D,k}(m)\mathcal{N}(\mathbf{x};\mathbf{x}^{i,j}_{k|k-1}(m),\mathbf{P}^{i,j}_{k|k-1}(m)) \nonumber \\
%  & \times 
%  \mathcal{N}(\mathbf{z};\mathbf{h}_k^{(m)}(\mathbf{x}),\bm{\Sigma}_{k}(\mathbf{z}))\nonumber \\
%  & \approx 
% \gamma^{i,j}_{k|k-1}(m)p^{i,j}_{D,k}(m)\mathcal{N}(\mathbf{x};\mathbf{x}^{i,j}_{k|k}(m),\mathbf{P}^{i,j}_{k|k}(m)) \nonumber \\
%  & \times 
%  \mathcal{N}(\mathbf{z};\mathbf{h}_k^{(m)}(\mathbf{x}^{i,j}_{k|k-1}(m)),\mathbf{S}_{\mathbf{z}\mathbf{z},k}^{i,j,(m)}), \label{eq:BayesUpdateApproximation}
%  %p^i_{\text{D}}(\mathbf{x},m)p(\mathbf{z}|\mathbf{x},\mathbf{s}^i_k,m)D^i_{k|k-1}(\mathbf{x},m). \label{eq:nu-function}
% \end{align}
We will denote $j(\mathbf{z})$ the birth index $j$ corresponding to measurement $\mathbf{z}$.  Considering a particular measurement $\mathbf{z}$, then when 
where for a birth  $j=j(\mathbf{z})$, \cite{Ristic2012}
  \begin{align}\label{eq:nu-birth}
      \nu^{i,j}(\mathbf{z},\mathbf{x},m)= \gamma^{i,j}_{k|k-1}(m)\mathcal{N}(\mathbf{x};\mathbf{x}^{i,j}_{k|k-1}(m),\mathbf{P}^{i,j}_{k|k-1}(m)), 
  \end{align}
  while for any $j\neq j(\mathbf{z})$
  \begin{align}\label{eq:BayesUpdateApproximation}
  \nu^{i,j}(\mathbf{z},\mathbf{x},m) &= p^{i,j}_{\text{D},k}(m)\, \gamma^{i,j}_{k|k-1}(m)\, \mathcal{N}(\mathbf{x};\mathbf{x}^{i,j}_{k|k-1}(m),\mathbf{P}^{i,j}_{k|k-1}(m)) 
  \, \mathcal{N}(\mathbf{z};\mathbf{h}^i(\mathbf{x},m),\bm{\Sigma}_{k}(\mathbf{z})) \\
 & \approx 
 p^{i,j}_{\text{D},k}(m)\, \gamma^{i,j}_{k|k-1}(m)\, \mathcal{N}(\mathbf{x};\mathbf{x}^{i,j}_{k|k}(m),\mathbf{P}^{i,j}_{k|k}(m))
 \, \mathcal{N}(\mathbf{z};\mathbf{h}^{i}(\mathbf{x}^{i,j}_{k|k-1}(m),m),\mathbf{S}_{\mathbf{z}\mathbf{z},k}^{i,j}(m)), \nonumber
\end{align}
where $\bm{\Sigma}_{k}(\mathbf{z})$ is the measurement covariance of measurement $\mathbf{z}$.
The approximation in \eqref{eq:BayesUpdateApproximation} follows from the CKF, described in CKF update of Algorithm \ref{alg:MMPHD_Correction_step1}, \ref{alg:MMPHD_Correction_step2} in Appendix \ref{sec:pseudoCodeMap}.

\subsubsection{Vehicle Update (\ref{eq:vehStateUpdate})}
    Computing \eqref{eq:likelihood} in the log-domain, log-weight update is\footnote{The logarithm term $\log ( c(\mathbf{z}) + \sum_{m'}\int \nu^{i}(\mathbf{z},\mathbf{x}',m')\text{d}\mathbf{x}')$ is implemented by first introducing $W^{i,j}(\mathbf{z})= c(\mathbf{z}) + \sum_{m'}\int \nu^{i,j}(\mathbf{z},\mathbf{x}',m')\text{d}\mathbf{x}'$,  sorting these (for a given $i$) from large to small and re-indexing. Then
    $\log ( c(\mathbf{z}) + \sum_{m'}\int \nu^{i}(\mathbf{z},\mathbf{x}',m')\text{d}\mathbf{x}')=\log W^{i,(1)} (\mathbf{z}) + 
    \log (1 + \sum_{l>1} W^{i,(l)}(\mathbf{z})/W^{i,(1)}(\mathbf{z}))$.} 
\begin{align}
    \ell_{k|k}^{i} = \ell_{k|k-1}^{i} + \sum_{\mathbf{z}\in\mathcal{Z}} \log ( c(\mathbf{z}) + \sum_{m'}\int \nu^i(\mathbf{z},\mathbf{x}',m')\text{d}\mathbf{x}').
\end{align}   
    {We note that the closed form evaluation in \eqref{eq:nu-sum}--\eqref{eq:BayesUpdateApproximation} is  used for evaluating \eqref{eq:likelihood}.}

    %\begin{align}
    %    & \log( c(\mathbf{z}) + \sum_{m'}\int \nu^i(\mathbf{z},\mathbf{x}',m')\text{d}\mathbf{x}' )  \\
     %   & = \log W^{i,(1)} (\mathbf{z}) + 
    %\log (1 + \sum_{l>1}W^{i,(l)}(\mathbf{z})/W^{i,(1)}(\mathbf{z})).\nonumber
    %\exp(\log W(\mathbf{z})^{i,(l)}-\log W(\mathbf{z})^{i,(1)}) ),
    %\end{align}
    Finally, we denote the estimated vehicle state and estimated vehicle location by $\hat{\mathbf{s}}_{k}$ and $\hat{\mathbf{v}}_{k}$, respectively.
    The vehicle state is estimated by {the sample mean,} $\hat{\mathbf{s}}_{k} = \sum_{i=1}^{I} \omega^i_{k|k}\mathbf{s}^i_{k}$, 
    and the estimated vehicle location $\hat{\mathbf{v}}_{k}\in \mathbb{R}^{3}$ is extracted from $\hat{\mathbf{s}}_{k}\in \mathbb{R}^{7}$.
    We denote the resampled particle set by $
    \{\bar{\mathbf{s}}^i_{k},\bar{\omega}^i_{k|k}\}_{i=1}^I$, $\bar{\omega}^i_{k|k} = 1/I~~\forall i$.

\section{Global processing: Map Fusion}
\label{sec:GlobalProcessingLKA}

In this section, we consider fusion of information from different vehicles. As mentioned in Section~\ref{sec:Introduction}, we aim to leverage the local processing capabilities of each vehicle, as described in Section \ref{sec:LocalProcessingMMPHD}. To allow simple processing, we consider the case where vehicles asynchronously communicate with the BS, where each communication involves an uplink transmission and a downlink transmission. Hence, a vehicle may only sporadically communicate with the BS. At the beginning of a time slot $k$, the BS maintains maps  $D^{\text{BS}}_{k|k-1}(\mathbf{x},m)$ in GM form, for  $m \in \{\text{SP},\text{VA}\}$.

\subsection{Uplink Transmission}\label{sec:UplinkTransmission}

At time $k$, a certain vehicle $n$ determines {a particle average PHD}
\begin{align}\label{eq:AverageMap}
    D^{(n)}_{k|k}(\mathbf{x},m) = \sum_{i=1}^{I} \omega_{k|k}^{i,(n)}D^{i,(n)}_{k|k}(\mathbf{x},m),
\end{align}
to which we apply pruning and merging\footnote{Gaussian components (mean, covariance, normalized weight) for all particles are imported as the input since our PHD uses the particle approach.} for implementation, described in  \cite[Table II]{Vo2006}. The vehicle sends the average PHD as well as a representation of the accumulated FoV since the last communication instant $k^\dagger$
\begin{align} \label{eq:FOVACC}
    \mathcal{F}^{(n)}_{k}(m)= \{ \mathbf{x}: \exists k'\in (k^\dagger,k], \max_i \{p_{\text{D},k'}(\mathbf{x},\hat{\mathbf{s}}^{i}_{k'},m)\} \ge \gamma_{\text{D}}\},
\end{align}
where $\gamma_{\text{D}}$ is a detection threshold (close to 1).  

\subsection{Map Fusion at the BS}
The BS receives $D^{(n)}_{k|k}(\mathbf{x},m)$ and fuses with the local map $D^{\text{BS}}_{k|k-1}(\mathbf{x},m)$. There are two common approaches for fusing two PHDs
$D_1 (\mathbf{x})$ and $D_2 (\mathbf{x})$
\begin{align}
    D_\beta (\mathbf{x}) =\begin{cases}
    (D_1 (\mathbf{x}))^{\beta_1} (D_2(\mathbf{x}))^{\beta_2} & \text{GCI~fusion~\cite{MahlerGCI}}\\
    {\beta_1}D_1(\mathbf{x}) + {\beta_2}D_2(\mathbf{x}) & \text{AA~fusion~\cite{li2017generalized,li2018local}}
    \end{cases}   
\end{align}
where $\beta_1$, $\beta_2$ are the fusion weights which satisfy
$\beta_1,\beta_2 \ge 0$ and $\beta_1+\beta_2 =1$. The values of $\beta_1$ and $\beta_2$ are set to reflect the relative contributions of $D_1$ and $D_2$.
From the information-theoretic point of view, both approaches lead to a fused
PHD that can be interpreted as the (respectively left- and right-) centroid of the PHDs to be fused when the Kullback-Leibler divergence is used as discrepancy measure \cite{Gao2019multiobject}. 
However, the two fusion rules have different characteristics. For instance, 
due to its multiplicative nature, GCI tends to
preserve only objects present in all the PHDs to be fused and, hence,  is preferable when the PHDs to be fused originate from sensors having a high clutter rate.
On the other hand, AA is more suitable for higher rates of missed detections since
it tends to preserve all the detected objects. 
Thus, it is clear that GCI fusion is hard to combine with sensors that have limited FoVs,
since, by definition, the probability of detection is equal to zero outside each vehicle FoV \cite{Mullane2011}.
% because the limited FoVs increase the chances of misdetected objects.  
For this reason, we choose to use AA.

Let $\mathcal I^{(n)}_k(m)$ denote the region of the map space where vehicle $n$ has information. Notice that such a region includes the accumulated FoV 
$ \mathcal{F}^{(n)}_{k}(m)$ but also all the regions containing the components of the map $D^{(n)}_{k|k}(\mathbf{x},m)$.
In fact, each vehicle can have information also outside its own FoV thanks to the downlink transmission 
from the BS to the vehicles. Similarly, let $\mathcal I^\text{BS}_k(m)$ denote the region of the map space where the BS has information. 
Accordingly,
the source sets ${\cal X}_k$ is divided into three mutually disjoint sets:
\begin{enumerate}[(a)]
    \item the set $\overline {\mathcal X}_k$ of sources  on which both the BS and vehicle $n$ have information, belonging to the intersection $\mathcal I^{(n)}_k(m) \cap \mathcal I^\text{BS}_k(m)$; \item the set $\widehat {\mathcal X}_k$ of sources 
on which only vehicle $n$ has information, belonging to the relative complement $\mathcal I^{(n)}_k(m) \setminus \mathcal I^\text{BS}_k(m)$; and 
    \item the set $\widetilde {\mathcal X}_k$ of sources
 on which only the BS has information, belonging to the relative complement
$\mathcal I^\text{BS}_k(m) \setminus  \mathcal I^{(n)}_k(m)$. 
\end{enumerate}
It is clear that the three sets $\overline {\mathcal X}_k$, $\widehat {\mathcal X}_k$, and $\widetilde {\mathcal X}_k$ 
have to be considered separately because an actual information fusion is possible only for the sources
belonging to  $\overline {\mathcal X}_k$. To this end,
we can exploit the property that the PHD of the union of independent Poisson RFSs is the sum of the PHDs and write
\begin{eqnarray}
D^{\text{BS}}_{k|k-1}(\mathbf{x},m) &=& \overline D^{\text{BS}}_{k|k-1}(\mathbf{x},m) + \widetilde D^{\text{BS}}_{k|k-1}(\mathbf{x},m) + \widehat D_{k|k - 1}^{BS}({\bf x},m), \label{eq:dec1} \\
D^{(n)}_{k|k}(\mathbf{x},m) &=& \overline D^{(n)}_{k|k}(\mathbf{x},m)  + \widetilde D_{k|k}^{(n)}({\bf x},m) + \widehat D^{(n)}_{k|k}(\mathbf{x},m), \label{eq:dec2}
\end{eqnarray}
where $\overline D^{\text{BS}}_{k|k-1}(\mathbf{x},m)$ and $\overline D^{(n)}_{k|k}(\mathbf{x},m)$ refer to $\overline {\mathcal X}_k$, 
$\widetilde D^{\text{BS}}_{k|k-1}(\mathbf{x},m)$ and $\widetilde D_{k|k}^{(n)}({\bf x},m)$ refer to $\widetilde {\mathcal X}_k$,
and finally $\widehat D_{k|k - 1}^{BS}({\bf x},m)$ and $\widehat D^{(n)}_{k|k}(\mathbf{x},m)$ refer to $\widehat {\mathcal X}_k$. 
As can be concluded immediately, by construction, we have $\widetilde D_{k|k}^{(n)}({\bf{x}},m) = 0$ and $\widehat D_{k|k - 1}^{{\rm{BS}}}({\bf{x}},m) = 0$.
Then the idea is to carry out the AA fusion independently on the three sets $\overline {\mathcal X}_k$, $\widehat {\mathcal X}_k$, and $\widetilde {\mathcal X}_k$ and, accordingly, the fusion result at the BS takes the form %$D_{k|k}^{{\rm{BS}}}({\bf{x}},m) = \overline D_{k|k}^{{\rm{BS}}}({\bf{x}},m) + \widetilde D_{k|k}^{{\rm{BS}}}({\bf{x}},m) + \widehat D_{k|k}^{{\rm{BS}}}({\bf{x}},m)$, 
\begin{align} \label{eq:fusion}
     D_{k|k}^{{\rm{BS}}}({\bf{x}},m) = \overline D_{k|k}^{{\rm{BS}}}({\bf{x}},m) + \widetilde D_{k|k}^{{\rm{BS}}}({\bf{x}},m) + \widehat D_{k|k}^{{\rm{BS}}}({\bf{x}},m),
\end{align}
where 
\begin{align}
    \overline D_{k|k}^{{\rm{BS}}}({\bf{x}},m) &= {{\overline \beta }_p}\overline D_{k|k - 1}^{{\rm{BS}}}({\bf{x}},m) + {{\overline \beta }_a}{\mkern 1mu} \overline D_{k|k}^{(n)}({\bf{x}},m), \label{eq:FC} \\
    \widetilde D_{k|k}^{{\rm{BS}}}({\bf{x}},m) &= {{\widetilde \beta }_p}\widetilde D_{k|k - 1}^{{\rm{BS}}}({\bf{x}},m) + {{\widetilde \beta }_a}\widetilde D_{k|k}^{(n)}({\bf{x}},m).  %= {{\widetilde \beta }_p}\widetilde D_{k|k - 1}^{{\rm{BS}}}({\bf{x}},m), 
    \label{eq:FBS} \\
    \widehat D_{k|k}^{{\rm{BS}}}({\bf{x}},m) &= {{\widehat \beta }_p}\widehat D_{k|k - 1}^{{\rm{BS}}}({\bf{x}},m) + {{\widehat \beta }_a}\widehat D_{k|k}^{(n)}({\bf{x}},m).% = {{\widehat \beta }_a} \widehat D_{k|k}^{(n)}({\bf{x}},m).   
    \label{eq:Fn}
\end{align}
Here different fusion weights have been assigned for the three disjoint sets since each fusion is supposed to be 
independently carried out. Such an additional flexibility allows us to take into account directly in the fusion rule
the decomposition of the source sets.
%the decompositions at (\ref{eq:dec1}) and (\ref{eq:dec2}) are supposed to be independently carried out.
Due to the fact that both the BS and the vehicle have information on the common set $\overline {\cal X}_k$,
the uniform weights ${{\overline \beta }_p} = {{\overline \beta }_a} = 1/2$ can be adopted for $\overline {\cal X}_k$, while for $\widetilde  {\cal X}_k$ (the source set for which the vehicle has no information) the weight can be set to ${{\widetilde \beta }_a}=0$, ${{\widetilde \beta }_p}=1$, and for $\widehat  {\cal X}_k$ (the source set for which the BS has no information) the weight can be set to ${{\widehat \beta }_a}=1$, ${{\widehat \beta }_p}=0$.\footnote{Setting ${{\widehat \beta }_a} = 1/2$ would lead to a reduction by a half of the weights of the Gaussian
components related to the sources belonging to ${\widehat {\cal X}_k}$, which hence would become
lower than the threshold for declaring the presence of the source (i.e. $0.5$).
Then,  if these sources are not detected again, 
the fused PHD will not be able to declare them anymore.
Such a situation can happen to newly detected landmarks located at the border of the FoV or when the vehicles move fast.
It has been pointed out in \cite{bunn1975bayesian} that
the fusion weight should be selected according to the probability
that the next prediction made using its corresponding density 
outperform predictions made from all other individual densities.
} 

In practice, the decompositions in (\ref{eq:dec1}) and (\ref{eq:dec2}) are not known and have to be
approximately determined directly form the densities $D_{k|k}^{(n)}(\mathbf{x},m)$ and ${D}^{\text{BS}}_{k|k-1}(\mathbf{x},m)$.
We conclude this section by presenting a procedure for deriving such decompositions
when the
average vehicle map and the BS map are  expressed as GMs
    \begin{align}
        D_{k|k}^{(n)}(\mathbf{x},m) = \sum_{j_a=1}^{\hat{J}_{k}}\hat{\gamma}_{k}^{j_a}\mathcal{N}(\mathbf{x};\hat{\mathbf{x}}_k^{j_a},\hat{\mathbf{P}}_k^{j_a}),~~
        {D}^{\text{BS}}_{k|k-1}(\mathbf{x},m) = \sum_{j_p=1}^{\hat{J}_{p,k}}\hat{\gamma}_{k}^{j_p}\mathcal{N}(\mathbf{x};\hat{\mathbf{x}}_k^{j_p},\hat{\mathbf{P}}_k^{j_p}).
    \end{align}
We notice preliminarily that, in this case, the fusion rule
(\ref{eq:fusion}) can be rewritten as
\begin{eqnarray}
\label{eq:fusion2}
    D^{\text{BS}}_{k|k}(\mathbf{x},m) &=& \sum_{j_a=1}^{\hat{J}_{k}}
    \beta_{a,j_a} \, \hat{\gamma}_{k}^{j_a} \, \mathcal{N}(\mathbf{x};\hat{\mathbf{x}}_k^{j_a},\hat{\mathbf{P}}_k^{j_a} ) + \sum_{j_p=1}^{\hat{J}_{p,k}} \beta_{p,j_p} \, \hat{\gamma}_{k}^{j_p} \, \mathcal{N}(\mathbf{x};\hat{\mathbf{x}}_k^{j_p},\hat{\mathbf{P}}_k^{j_p}),
\end{eqnarray}
where $\beta_{a,j_a}$ takes value $1/2$ when the component $j_a$ is assigned
to $\overline D^{(n)}_{k|k}(\mathbf{x},m)$ or value $1$ when it is assigned to $\widehat D^{(n)}_{k|k}(\mathbf{x},m)$ and, similarly, $\beta_{p,j_p}$ takes value 
$1/2$ when the component $j_p$ is assigned
to
$\overline D^{\text{BS}}_{k|k-1}(\mathbf{x},m)$ or value $1$ when it is assigned to $\widetilde D^{\text{BS}}_{k|k-1}(\mathbf{x},m)$. 
       
To set the values of $\beta_{a,j_a}$ and $\beta_{p,j_p}$ (i.e. to approximately
determine the decompositions in (\ref{eq:dec1}) and (\ref{eq:dec2})), we use the Mahalanobis cost metric to compute the distance between the components of the two PHDs. 
%according to the \textcolor{olive}{GCI fusion rule} in \cite{li2019distributed}. This avoids double counting of information and operates as follow:  
%\begin{align}
%    D^{\text{BS}}_{k|k}(\mathbf{x},m) = \left(D^{\text{BS}}_{k|k-1}(\mathbf{x},m)\right)^{\beta_p}\left(D^{(n)}_{k|k}(\mathbf{x},m)\right)^{\beta_a},
%\end{align}
 % When the BS map is empty, we set $\beta_p=0$ and $\beta_a=1$. 
%An alternate approach is to perform AA fusion \textcolor{olive}{\cite{li2017generalized,li2018local}}:
%\begin{align}
%    D^{\text{BS}}_{k|k}(\mathbf{x},m) = {\beta_p}D^{\text{BS}}_{k-1|k}(\mathbf{x},m) + {\beta_a}D^{(n)}_{k|k}(\mathbf{x},m),
%\end{align} 
% \HW{@Giorgio: can you add some theoretical justification?}
%However, both GCI and AA require modifications to \textcolor{magenta}{work well} when \textcolor{magenta}{the vehicles} have different FoV \cite{Li2018MSMOFOV,li2019distributed}. Hence, AA and GCI can be applied for VA map fusion, but not for SP map fusion. For the VA map, we use AA with $\beta_p=\beta_a=1/2$. 
%For the SP map, we resort to the following simple approach to account for \textcolor{magenta}{the different FoVs}. We will drop the superscript $m=\text{SP}$ to simplify the notation. 
Specifically, we introduce two distance metrics
    \begin{align}
        \mathcal{D}_{j_a\rightarrow j_p} &  = \big(\hat{\mathbf{x}}_k^{j_a}-\hat{\mathbf{x}}_k^{j_p} \big)^\mathrm{T}({\hat{\mathbf{P}}_k^{j_a}})^{-1}\big(\hat{\mathbf{x}}_k^{j_a}-\hat{\mathbf{x}}_k^{j_p} \big),~\mathcal{D}_{j_p\rightarrow j_a} & = \big(\hat{\mathbf{x}}_k^{j_a}-\hat{\mathbf{x}}_k^{j_p} \big)^\mathrm{T}({\hat{\mathbf{P}}_k^{j_p}})^{-1}\big(\hat{\mathbf{x}}_k^{j_a}-\hat{\mathbf{x}}_k^{j_p} \big).
    \end{align}
    where $\mathcal{D}_{j_a\rightarrow j_p}$ is the Mahalanobis distance between $\hat{\mathbf{x}}_k^{j_p}$ and the distribution $\mathcal{N}(\mathbf{x};\hat{\mathbf{x}}_k^{j_a},\hat{\mathbf{P}}_k^{j_a})$ while $\mathcal{D}_{j_p\rightarrow j_a}$ is the Mahalanobis distance between $\hat{\mathbf{x}}_k^{j_a}$ and  $\mathcal{N}(\mathbf{x};\hat{\mathbf{x}}_k^{j_p},\hat{\mathbf{P}}_k^{j_p})$. With these metrics, we compute binary proximity matrices $\mathbf{C}^{a}\in \mathbb{B}^{\hat{J}_k} \times \mathbb{B}^{\hat{J}_{p,k}}$ and $\mathbf{C}^{p}\in \mathbb{B}^{\hat{J}_k} \times \mathbb{B}^{\hat{J}_{p,k}}$, initialized as zeros. Then, we cycle through all pairs $(j_a,j_p)$: if $\mathcal{D}_{j_p\rightarrow j_a}<\gamma_\text{UP} $, then we set ${C}^{p}_{j_a,j_p}=1$. If $\mathcal{D}_{j_a\rightarrow j_p}<\gamma_\text{UP}$, then we set ${C}^{a}_{j_a,j_p}=1$. Here, $\gamma_\text{UP}$  is a threshold on the Mahalanobis distances.
Finally, we determine the $\beta$ values for each component. We initialize ${D}^{\text{BS}}_{k|k}(\mathbf{x},m)=0$ and add entries as follows:
    \begin{enumerate}
     \item \emph{Assign equal weights for matches:}
    If $\max({C}^{p}_{j_a,j_p},{C}^{a}_{j_a,j_p})=1$,
    % \LG{which indicates Gaussian components $j_a$ of $D_{k|k}^{(n)}({\bf{x}},m)$ and $j_p$ of $D_{k|k - 1}^{{\rm{BS}}}({\bf{x}},m)$ describe the same feature,}
    the components $j_a$ of $D_{k|k}^{(n)}({\bf{x}},m)$ and $j_p$ of $D_{k|k - 1}^{{\rm{BS}}}({\bf{x}},m)$ are deemed to belong to the region of common information $\mathcal I^{(n)}_k(m) \cap \mathcal I^\text{BS}_k(m)$, 
    since both of them can find their respective correspondences in the other map PHDs. Thus
    we set  $\beta_{p,j_p}=\beta_{a,j_a}=1/2$. Note that a source $j_p$ could be matched with multiple sources $j_a$ and vice versa. 
    \item \emph{Find unmatched sources in the BS map:}
    If $\sum_{j_a}{C}^{p}_{j_a,j_p}=0$, then source $j_p$ in the BS map could not be associated with any entry in the vehicle map. Then,
    recalling that $\mathcal I^{(n)}_k(m)$ always contains the accumulated FoV $\mathcal{F}^{(n)}_{k}(m)$,
    we add the source $j_p$ to ${D}^{\text{BS}}_{k|k}(\mathbf{x},m)$ with weight
    \begin{align}
    \beta_{p,j_p}=
        \begin{cases}
        1 & \hat{\mathbf{x}}_k^{j_p}\notin\mathcal{F}^{(n)}_{k}(m),\\
        1/2 & \hat{\mathbf{x}}_k^{j_p}\in\mathcal{F}^{(n)}_{k}(m).
        \end{cases}
    \end{align} 
    This ensures that sources outside the FoV are kept.
    However, sources {that} suddenly appear {could possibly be false alarms}, therefore sources in the field of view that were not seen by vehicle $n$ are reduced in weight and will gradually disappear from the BS map.
    \item \emph{Find unmatched sources in the vehicle map:}
    If $\sum_{j_p}{C}^{p}_{j_a,j_p}=0$, then  source $j_a$ in the vehicle map could not be associated with any entry in the BS map. Hence, 
    the component  $j_a$ is deemed to belong
    to the relative complement $\mathcal I^{(n)}_k(m) \setminus \mathcal I^\text{BS}_k(m)$ and we set $\beta_{a,j_a}=1$. 

    \end{enumerate}
The BS map is then found by adding all these 
sources with their corresponding weights as in (\ref{eq:fusion2}) and
by applying pruning and merging so as to keep the number of components limited. 
Clearly, at the beginning, when the
BS map is empty, instead of applying (\ref{eq:fusion2}) the BS map is simply overwritten with the vehicle map. 

    \subsection{Downlink Transmission} \label{sec:downlinkTransmission}
    The BS sends the computed ${D}^{\text{BS}}_{k|k}(\mathbf{x},m)$ to vehicle $n$. This map can contain new information for the vehicle as it contains all the information provided by \emph{other} vehicles between times $k^\dagger$ and $k$. Hence,
    %In the down-link stage, the map fusion is simply implemented compared to the up-link stage.
%When the vehicle receives the fused maps at time $k+1$, for the map prediction (i.e., \eqref{eq:MapPredictionModel}), 
the vehicle overwrites the fused map to $D^i_{k|k}(\mathbf{x},m)$ as follows
\begin{align}
    D^i_{k|k}(\mathbf{x},m) = D^{\text{BS}}_{k|k}(\mathbf{x},m) ~~ \forall i.
\end{align}
While this leads to a lack of diversity among the maps across the particles, it has the distinct benefit of being a low-complexity solution.

\section{Numerical Results}\label{sec:NumericalResults}
% \HW{I propose to have a scenario with two vehicles driving towards each-other on a two-lane road. Then each vehicle will map part of the environment and (later) benefit from what the other vehicle has mapped.}

\begin{figure}
\begin{centering}
	{\includegraphics[width=.4\columnwidth]{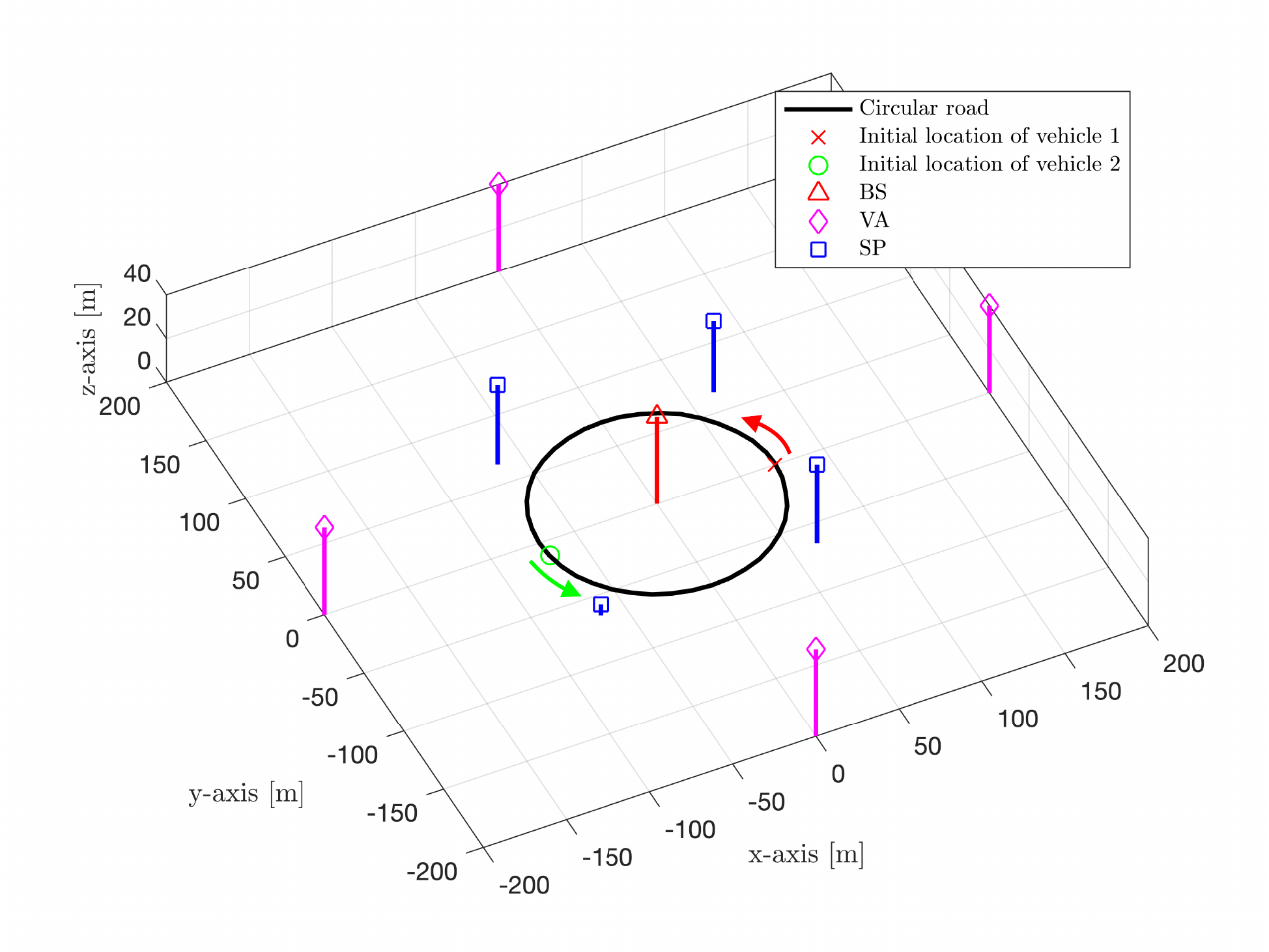}}
	\caption{Vehicle trajectories and common map of the environment (1 BS, 4 VAs, and 4 SPs). Two vehicles are initially located at opposite sides of $[0,0,0]~\mathrm{m}$, and move along with counterclockwise. Two vehicles have observed same objects (i.e., explored area) when $k$ is about 40.}
	\label{Fig:VehicularNetwork}
	\par
\end{centering}
\end{figure}

\subsection{Simulation Setup}\label{sec:Setup}

     We consider a three dimensional (3D) vehicular network where two moving vehicles are on a circular road with a BS, four VAs, and four SPs as shown in Fig. \ref{Fig:VehicularNetwork}.
    Such a scenario is sufficiently representative to demonstrate the  efficacy of the proposed PHD filter and map fusion, though actual performance in localization and mapping will depend on the chosen scenario. The details of the scenario are shown and available in \cite{Hyowon_5GCPM_Dataport}.
    %we set that two vehicles have observed same explored area after each vehicle is finished one-round moving.}
    During $K=40$ time steps, the vehicle states are evolved with the dynamics model  \eqref{eq:processModel} as discussed in Section \ref{sec:VehicleEnvironmentModel}, with 
\begin{align}\label{eq:processModel2}
\bm{\upsilon}(\mathbf{s}^{(n)}_{k-1}) =\mathbf{s}^{(n)}_{k-1}+
\left[\begin{array}{c}
\frac{\zeta^{(n)}_{k-1}}{\rho^{(n)}_{k-1}}\left(\sin(\alpha^{(n)}_{k-1}+\rho^{(n)}_{k-1}\Delta)-\sin\alpha^{(n)}_{k-1}\right)\\
\frac{\zeta^{(n)}_{k-1}}{\rho^{(n)}_{k-1}}\left(-\cos(\alpha^{(n)}_{k-1}+\rho^{(n)}_{k-1}\Delta)+\cos\alpha^{(n)}_{k-1}\right)\\
0\\
 \rho^{(n)}_{k-1}\Delta\\
\mathbf{0}_3
\end{array}\right],
\end{align}
where $\mathbf{0}_\psi$ denotes a column vector of $\psi$ zeros, $\Delta$ is the sampling time and  $\mathbf{q}_{k}$ denotes the process noise, modeled as zero-mean Gaussian with covariance  $\mathbf{Q}=\mathrm{diag}[\sigma_x^2,\sigma_y^2,0,\sigma^2_{\alpha} ,\mathbf{0}_2,\sigma^2_B]$.
The vehicles are initialized as $\mathbf{s}^{(1)}_0 = [70.7285~\text{m}, 0~\text{m}, 0~\text{m}, \pi/2~\text{rad}, 22.22~\text{m/s}, \pi/10~\text{rad/s}, 300~\text{m}]^\text{T}$ and $\splitatcommas{\mathbf{s}^{(2)}_0 = [-70.7285~\text{m}, 0~\text{m}, 0~\text{m},\pi/2~\text{rad}, -22.22~\text{m/s}, \pi/10~\text{rad/s}, 300~\text{m}]^\text{T}}$.
% , with units m, m, m, rad, m/s, rad/s, and m.
    The time interval $\Delta$ is set to $0.5$ s.
    The process noise standard deviations are set to $\sigma_x=0.2$ m, $\sigma_y=0.2$ m, $\sigma_{\alpha}=0.001$ rad, and $\sigma_B=0.2$ m.
    The initial prior of the vehicle state follows zero-mean Gaussian distribution with $0.3$ m standard deviation for both x and y location, $0.3$ rad for the vehicle heading, $0.3$ m the bias.
    The longitudinal velocity $\zeta^{(n)}_{k}$, rotational velocity $\rho^{(n)}_{k}$ are assumed to be known.
    The measurement covariance matrix $\bm{\Sigma}^{(n)}_{k,l}$ is diagonal, and is set to $\text{diag} (10^{-2}~\text{m}^2, 10^{-4}~\text{rad}^2, 10^{-4}~\text{rad}^2, 10^{-4}~\text{rad}^2)$.
    % , with units $\text{m}^2$, $\text{rad}^2$, $\text{rad}^2$, $\text{rad}^2$, and $\text{rad}^2$.
    To mitigate the effect of the errors in the CKF (due to the non-invertible nonlinearity), we replace the measurement $\bm{\Sigma}^{(n)}_{k,l}$ with $\bm{\Sigma}_{\textrm{PHD}}= 9 \times \bm{\Sigma}^{(n)}_{k,l}$ in \eqref{eq:Factorize} for the birth process, and in the CKF update of Algorithm \ref{alg:MMPHD_Correction} for the map correction.
    A BS is located at $[0, 0, 40]^\text{T}$ m. Four VAs are located at $[200, 0, 40]^\text{T}$ m, $[-200, 0, 40]^\text{T}$ m, $[0, 200, 40]^\text{T}$ m, $[0, -200, 40]^\text{T}$ m.
    Four SPs are located at $[65, 65, z_\text{SP}]^\text{T}$ m, $[-65, 65, z_\text{SP}]^\text{T}$ m, $[-65, -65, z_\text{SP}]^\text{T}$ m, and $[65, -65, z_\text{SP}]^\text{T}$ m, where $z_\text{SP}\sim \mathcal{U}(0,40)$.
    The SPs are only visible when the distance between the SP and vehicle is within the FoV  range $r_\text{FoV}=50~\text{m}$, while VAs are always visible. The detection probability $p_{\text{D}}$ is set to 0.9 within the FOV.  %which is set to $50$ m, and we set the uncertainty $r_\text{UC}$ to $3$ m.
    In the Gaussian representation of the birth process \eqref{eq:BirthProcessGM}, we consider the birth weight $\gamma_{b,k}^{i,j(\mathbf{z})}(m)=1.5\times10^{-5}$ for $m=\{\textrm{VA,SP}\}$. For the clutter intensity $c(\mathbf{z})$, we consider the average of the number of clutter measurements (following Poisson distribution) $\lambda=1$, and the maximum sensing range $R_{\max}=200$ m, so $c(\mathbf{z})  = \lambda/(4R_{\max}\pi^4)$.
    We utilized the pruning and merging in \cite[Table II]{Vo2006}, and also used its parameter notations as follows: truncation threshold $T$; merging threshold $U$; and maximum allowable number of Gaussians $J_{\max}$. We considered $T = 10^{-4}$, $U = 49$, and $J_{\max} = 50$. The object detection parameters are set to as follows: the VA detection threshold $T_\text{VA} = 0.7$; the SP detection threshold $T_\text{SP} = 0.55$.
    We consider an asynchronous map fusion where each vehicle communicates with the BS every 4 time steps, with vehicle 1 starting at time 10 and vehicle 2 at time 12. 
   % with the time scheduling where vehicle 1 transmits its particle average map PHD at time $K_T^{(1)} =\{10,14,...,38\}$, and at time $K_T^{(2)}=\{12,16,...,40\}$, so do vehicle 2. Then, the last time vehicle sent uplink map information $k^\dagger$ for vehicle 1 and 2 is addressed by $K_T^{(1)}$ and $K_T^{(2)}$, respectively.
    Each vehicle's state was represented by $I=2000$ particles, and simulation results were obtained by averaging over $N_\mathrm{mc}=20$ Monte Carlo runs (we observed no significant performance differences when increasing $N_\mathrm{mc}$ from 10 to 20).
    Complete source code is available at
    \url{https://github.com/HyowonKim-P1/5GmmWavePHDFilterMapFusion}.
    
\begin{figure}[t!]
    \begin{centering}
        \includegraphics[width=0.5\columnwidth]{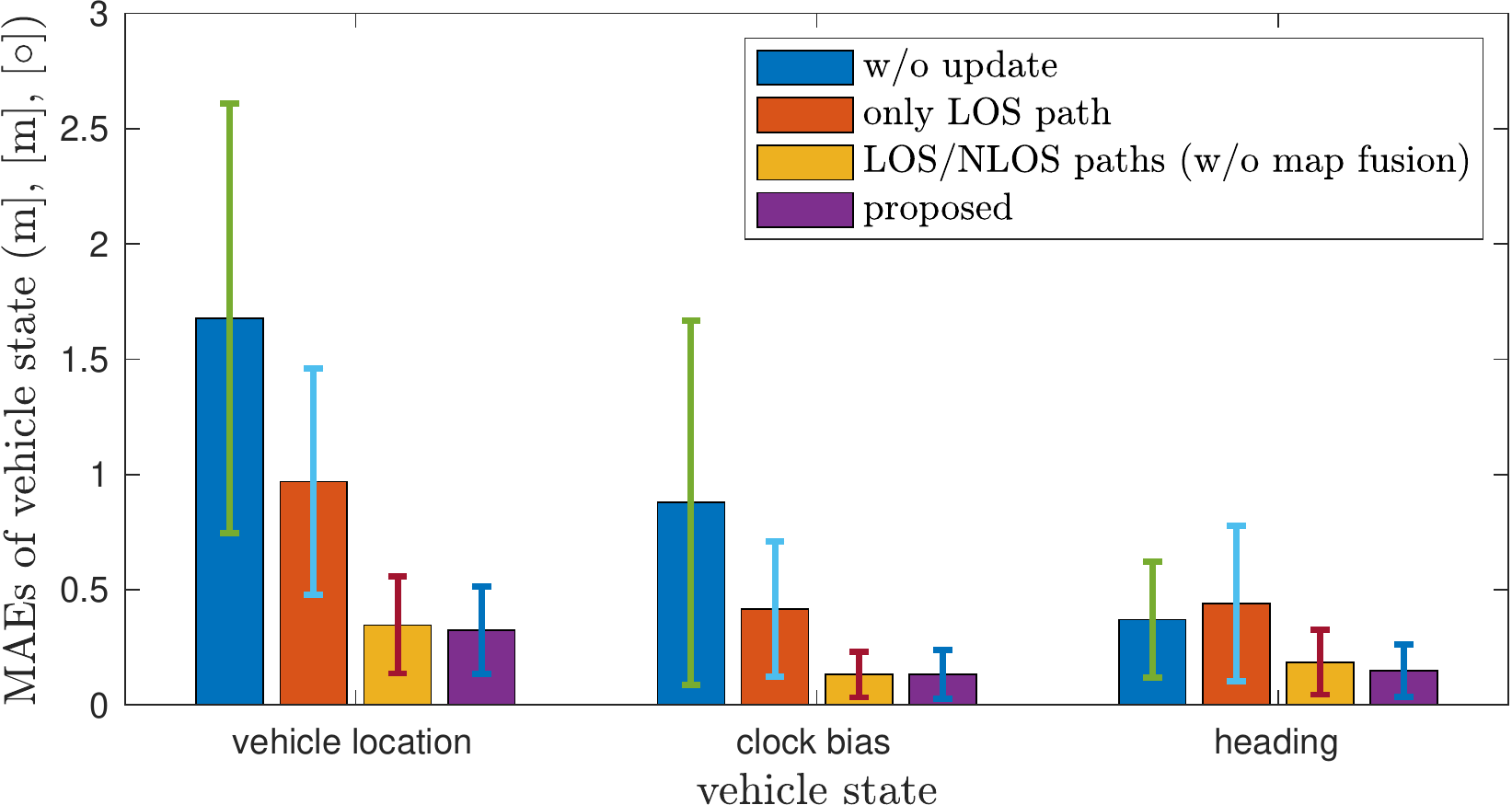}
    	\caption{MAEs and RMSE bars of the vehicle states (vehicle location, clock bias, and heading).}
    	\label{Fig:MAERMSEbar}
    	\par
    \end{centering}
    \end{figure}
    
\subsection{Performance Metric}
To demonstrate the efficacy of the method and support the contributions of this paper described in Section \ref{sec:Introduction}.
The performances of the vehicle state estimation and the mapping of the environment are evaluated, over all Monte Carlo runs during the  steady-state operation, which was determined to be after $k=20$. 
For the vehicle state estimation, we compute the mean absolute error (MAE) on each component (location, clock bias, heading), along with 
root mean square error (RMSE) bars. 
For the mapping, we compute the average of the generalized optimal subpattern assignment (GOSPA) distance \cite{RahmathullahGFS:2017}, as follows (removing the time index $k$ and the source type index $m$). We denote $\mathcal{L} = \{(\mathbf{x}_{1},m_1),\ldots,(\mathbf{x}_{M},m_{M})\}$ and $\hat{\mathcal{L}} = \{(\hat{\mathbf{x}}_{1},\hat{m}_1),\ldots,(\hat{\mathbf{x}}_{\hat{M}},\hat{m}_{\hat{M}})\}$ by the set of sources (of type $m$ at time $k$) and its estimated set, respectively. The FoV was not considered in the GOSPA distance metric in order to evaluate  the map fusion performance.  Then, the GOSPA is defined as
    \begin{align}
        d^{(q_c,q_a,q_p)}_{\text{GOSPA}}
        = \bigg(\min\limits_{\mathcal{M}\in \Pi_{|\hat{\mathcal{L}}|}} \sum_{e=1}^{M}
        d^{q_c}(\hat{\mathbf{x}}_e,\hat{\mathbf{x}}_{\mathcal{M}(e)})
        + \frac{q_c^{q_p}}{q_a}(\hat{M}-M) \bigg)^{\frac{1}{q_p}},
    \end{align}
    where $\Pi_{\star}$ indicates the permutations of set $\star$, cut off distance $q_c=20$, $q_a = 2$, power parameter $q_p=2$, and $d^{q_c}(\mathbf{x}_e,\hat{\mathbf{x}}_{\mathcal{M}(e)}) = \min(\lVert \hat{\mathbf{x}_e}-\hat{\mathbf{x}}_{\mathcal{M}(e)} \rVert, q_c)$.
    
    \subsection{Results and Discussions}
    
    \begin{figure}[t!]
\begin{centering}
	\subfloat[\label{Fig:GOSPA_VA_UL}]
	{\includegraphics[width=0.5\columnwidth]{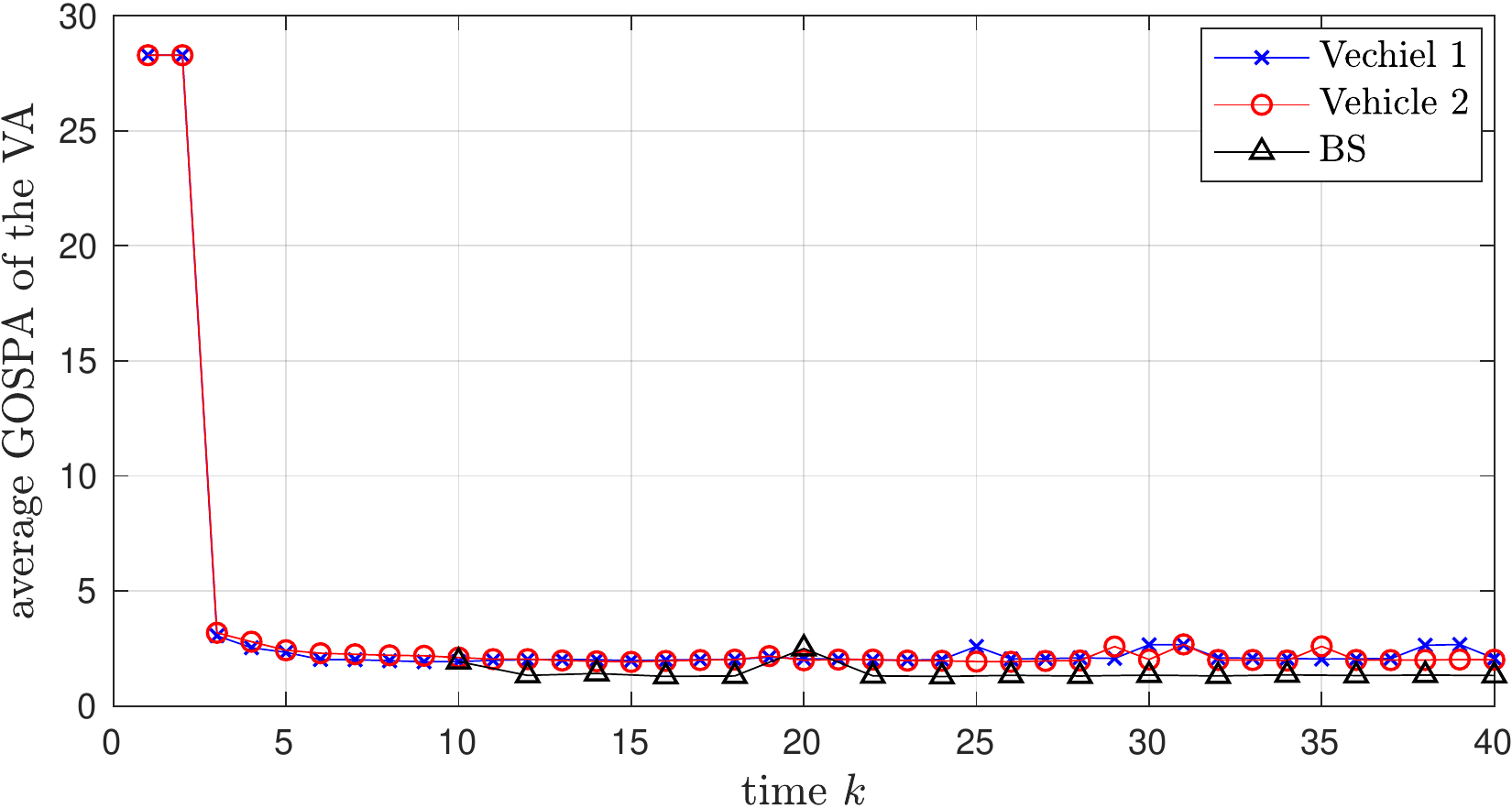}}\hfill
	\subfloat[\label{Fig:GOSPA_VA_UDL}]
	{\includegraphics[width=0.5\columnwidth]{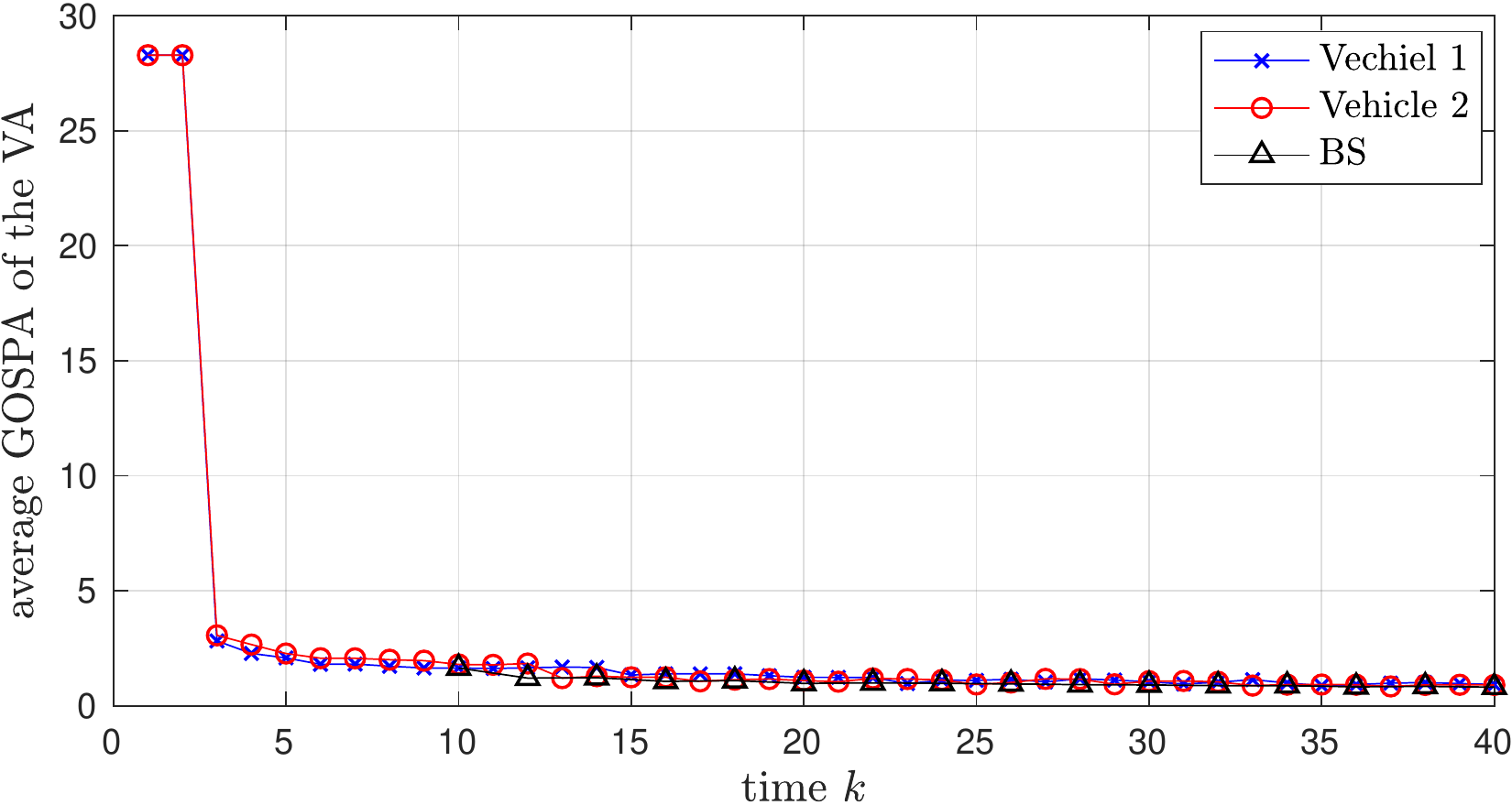}}
	\caption{Average GOSPA of the VA for (a) map fusion per vehicle and uplink transmission to the BS (b) map fusion per vehicle and uplink transmission to the BS as well as downlink transmission to the vehicles.}
	\label{Fig:GOSPA_VA}
	\par
\end{centering}
\end{figure}

\begin{figure}
\begin{centering}
	\subfloat[\label{Fig:GOSPA_SP_UL}]
	{\includegraphics[width=0.5\columnwidth]{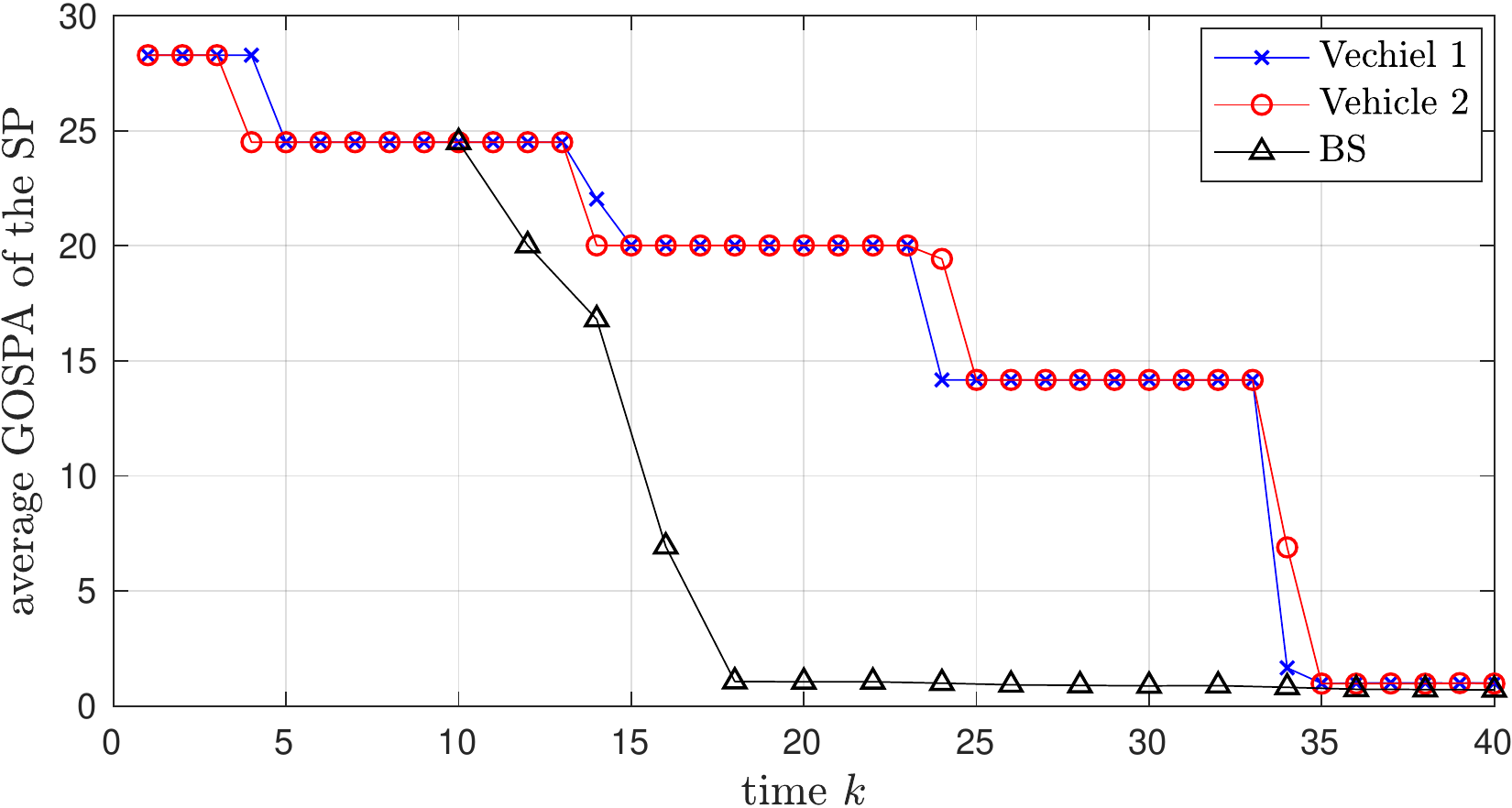}}\hfill
	\subfloat[\label{Fig:GOSPA_SP_UDL}]
	{\includegraphics[width=0.5\columnwidth]{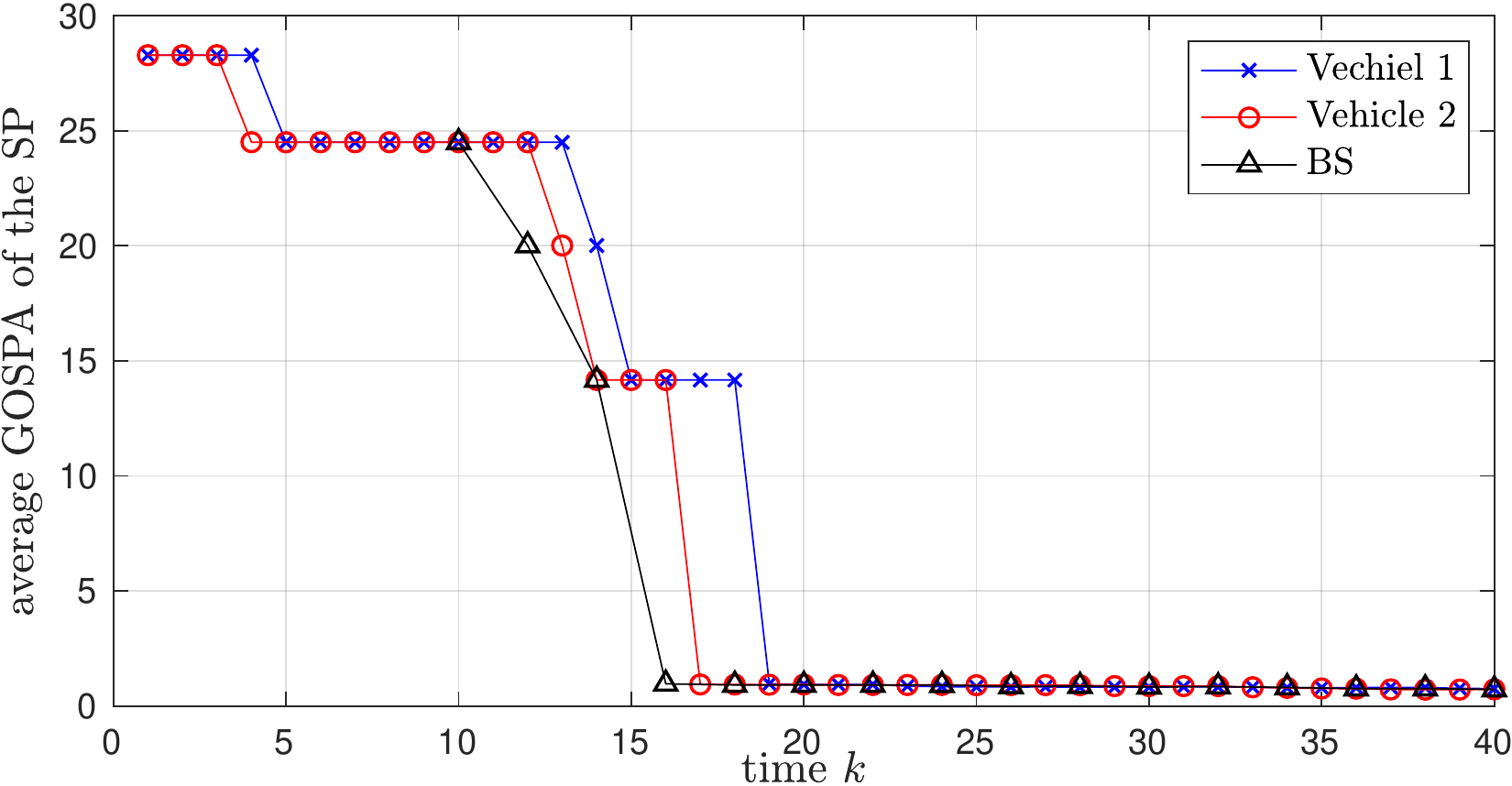}}
	\caption{Average GOSPA of the SP for (a) map fusion per vehicle and uplink transmission to the BS (b) map fusion per vehicle and uplink transmission to the BS as well as downlink transmission to the vehicles.}
	\label{Fig:GOSPA_SP}
	\par
\end{centering}
\end{figure}

    \subsubsection{Vehicle tracking}
     Fig. \ref{Fig:MAERMSEbar} shows the MAE and RMSE bars of the estimated vehicle location, clock bias, and heading with respect to (w.r.t.) the four cases as follows:
    %  case i) only performing the vehicle state prediction without the update step; case ii) using only the measurement from the LOS path; case iii) proposed PHD filter for positioning and mapping per vehicle  from Section \ref{sec:LocalProcessingMMPHD}; case iv) proposed PHD filtering and map fusion from Section \ref{sec:GlobalProcessingLKA}.
    i) only performing the vehicle state prediction without the update step; ii) using only the measurement from the LOS path; iii) proposed PHD filter for positioning and mapping per vehicle  from Section \ref{sec:LocalProcessingMMPHD}; and iv) proposed PHD filtering and map fusion from Section \ref{sec:GlobalProcessingLKA}.
    %\begin{itemize}
     %   \item[i)] only performing the vehicle state prediction without the update step;
      %  \item[ii)] using only the measurement from the LOS path;
       % \item[iii)] proposed PHD filter for positioning and mapping per vehicle  from Section \ref{sec:LocalProcessingMMPHD}; and
        %\item[iv)] proposed PHD filtering and map fusion from Section %\ref{sec:GlobalProcessingLKA}.
    %\end{itemize}
    In case i), the accuracy of the estimated vehicle state gradually increases and demonstrates the need for measurements in the considered scenario. 
    Case ii) can be considered a best-case, without any objects in the environment and a clear LOS at all times. We see that the performance is significantly improved compared to case i).  In case iii), the performance is much better than case ii) showing the benefit of NLOS information, even with unknown source association. In case iv), despite the reduced map diversity (see Section \ref{sec:downlinkTransmission}) the performance is not reduced
    % negatively affected
    compared to case iii), but there are only marginal performance gains either.
    % This is due to the specific scenario: for instance, a scenario with a second vehicle entering the environment after $k=40$ would have clear benefit from the BS map, built by the first vehicle. 
    This is due to the specific scenario, where the vehicles independently are able to localize themselves well. In addition, since the VAs are always visible for both vehicles and can be mapped accurately, the main cooperative localization gain can come from the SPs. Since SPs have variable detection probability, they provide only limited  information for the vehicles' positions.
%    The estimated propagation environment consisting of the objects is a statistical map as viewed by the vehicle state.
 %   Since the vehicles already have shared the map information to the propagation environment, it is hard to obtain the performance gain.
  %  However, a third vehicle entering the environment after $k=40$ could exploits the additional information to the propagation environment from the BS map, built by the both first and second vehicles.
   % Then, the third vehicle cooperatively perform 5G SLAM, and would have clear benefit in both positioning and mapping.}
    
    \subsubsection{Mapping}

    Fig. \ref{Fig:GOSPA_VA} shows the average GOSPA of the VA map, with  Fig.~\ref{Fig:GOSPA_VA_UL} considering the case of the local PHD filter and map fusion at the BS, but no downlink transmission, while  Fig.~\ref{Fig:GOSPA_VA_UDL} presents the performance of the proposed PHD filter and map fusion with downlink transmission. Comparing Fig.~\ref{Fig:GOSPA_VA_UL} and Fig.~\ref{Fig:GOSPA_VA_UDL}, there is only a small, little or no benefit the downlink transmissions, as both vehicles have all VAs in their FoV at all time. This is in contrast to the SP map, where Fig.~\ref{Fig:GOSPA_SP} reports the GOSPA results. In Fig.~\ref{Fig:GOSPA_SP_UL}, we see that the GOSPA per vehicle goes down as they move in the environment. The GOSPA at the BS is reduced faster, as it can benefit from the information of all vehicles. In Fig.~\ref{Fig:GOSPA_SP_UDL}, we note that when the BS sends back the map to the vehicles over the downlink, each vehicle can benefit from the measurements of the other vehicle, so that the GOSPA is reduced faster on the vehicle maps as well.

\section{Conclusions}\label{sec:Conclusions}
    It is expected that the framework of cooperative positioning and mapping with the proposed multiple-model PHD filter and map fusion is essential  \Ra{for vehicular positioning}.
    In this paper, we proposed a multiple-model PHD filter and map fusion for cooperative positioning and mapping in  vehicular networks \Ra{with 5G mmWave communication links}. The environment comprises a single BS, multiple vehicles, and different types of objects (small scattering objects and large reflecting surfaces).
    The challenges of the mapping such as the number of objects, object type, and their position were dealt with the proposed PHD filter.
    In addition, asynchronous map transmission to the BS is solved by the proposed map fusion method.
    From the results, it is confirmed that our PHD filter can handle the challenges of the mapping and vehicle state estimation simultaneously.
    We also confirmed that the proposed map fusion using map information of other vehicles significantly improves the mapping performance.

% if have a single appendix:
%\appendix[Proof of the Zonklar Equations]
% or
%\appendix  % for no appendix heading
% do not use \section anymore after \appendix, only \section*
% is possibly needed

% use appendices with more than one appendix
% then use \section to start each appendix
% you must declare a \section before using any
% \subsection or using \label (\appendices by itself
% starts a section numbered zero.)
%

\appendices

\section{Geometric Relations} \label{sec:Geometry}
In the relation between the observations (\ref{eq:ObservationModel2}), the state of the vehicle and the map depend on the origin of the measurement.
We distinguish between 3 different cases.

\subsection{Source is the BS}
For the LOS path between BS and vehicle, we have the following relations:  $\tau=\Vert\mathbf{x}_{\text{BS}}-\mathbf{v}_{k}\Vert/c+B$, where $c$ denotes the speed of light; $\phi_{0}^{\textrm{az}}=\arctan\left({y_{k}}/{x_{k}}\right)$,
$\phi_{0}^{\textrm{el}}=\arcsin\left(({z_{k}-z_{\textrm{BS}}})/({\|\mathbf{v}_{k}-\mathbf{x}_{\mathrm{BS}}\|})\right)$,
where we assume $\operatorname{arctan2}$ is used;  We remind that the DOA is measured in
the local frame of reference of the vehicle, so that the vehicle orientation
must be accounted for: $\theta_{0}^{\textrm{az}}=\pi+\arctan\left({y_{k}}/{x_{k}}\right)-\alpha_{k}$,
$\theta_{0}^{\textrm{el}}=\arcsin\left(({z_{\textrm{BS}}-z_{k}})/({\|\mathbf{x}_{\mathrm{BS}}-\mathbf{v}_{k}\|})\right)$,
since the DOA elevation measurement does not depend on the vehicle orientation. 

\subsection{Source is a reflecting surface}
Each reflecting surface can be parameterized by a fixed virtual anchor
(VA) location $\mathbf{x}_{\mathrm{VA}}$, obtained by mirroring the BS with respect to the surface. Between a virtual anchor $\mathbf{x}_{\text{VA}}$
and the user's position $\mathbf{v}_{k}$, the incidence point of
the specular reflection on the reflecting surface is given by the
point where the straight line between the VA and vehicle crosses the reflecting
surface
\begin{equation}
\mathbf{x}_{s}=\mathbf{x}_{\text{VA}}+\frac{(\mathbf{f}-\mathbf{x}_{\text{VA}})^{\mathrm{T}}\mathbf{u}}{(\mathbf{v}_{k}-\mathbf{x}_{\text{VA}})^{\mathrm{T}}\mathbf{u}}(\mathbf{v}_{k}-\mathbf{x}_{\text{VA}}).\label{eq:IncidencePoint}
\end{equation}
Here, $\mathbf{u}=(\mathbf{x}_{\text{BS}}-\mathbf{x}_{\text{VA}})/\Vert(\mathbf{x}_{\text{BS}}-\mathbf{x}_{\text{VA}})\Vert$
and $\mathbf{f}=(\mathbf{x}_{\text{BS}}+\mathbf{x}_{\text{VA}})/2$.
Note that this allows to find explicit expressions of $\mathbf{x}_{s}$
that only depend on $\mathbf{x}_{\text{VA}}$, $\mathbf{x}_{\text{BS}}$,
and $\mathbf{v}_{k}$ (not shown). Conversely, the location of a VA
can be expressed as a function of the incidence point 
\begin{align}
\mathbf{x}_{\mathrm{VA}}=\mathbf{v}_{k}+(\Vert\mathbf{v}_{k}-\mathbf{x}_{s}\Vert+\Vert\mathbf{x}_{\mathrm{BS}}-\mathbf{x}_{s}\Vert)\frac{\mathbf{x}_{s}-\mathbf{v}_{k}}{\Vert\mathbf{x}_{s}-\mathbf{v}_{k}\Vert}.\label{eq:mapSPtoVA}
\end{align}
Next, we state the relations between the channel parameters $\tau$,
$\bm{\theta}=[\theta^{\text{el}},\theta^{\text{az}}]^{\mathrm{T}}$,
and $\bm{\phi}=[\phi^{\text{el}},\phi^{\text{az}}]^{\mathrm{T}}$
and the system state:  $\tau=\Vert\mathbf{x}_{\mathrm{VA}}-\mathbf{v}_{k}\Vert/c+B$.
This is equivalent to $\tau=\Vert\mathbf{x}_{\mathrm{BS}}-\mathbf{x}_{s}\Vert/c+\Vert\mathbf{x}_{s}-\mathbf{v}_{k}\Vert/c+B$;  $\phi^{\textrm{az}}=\arctan\left({y_{\textrm{s}}}/{x_{\textrm{s}}}\right)$
and $\phi^{\textrm{el}}=\arcsin\left(({z_{\textrm{s}}-z_{\textrm{BS}}})/({\|\mathbf{x}_{\textrm{s}}-\mathbf{x}_{\mathrm{BS}}\|})\right)$; and  $\theta^{\textrm{az}}=\arctan\left(({y_{\textrm{VA}}-y_{k}})/({x_{\textrm{VA}}-x_{k}})\right)-\alpha_{k}$
and $\theta^{\textrm{el}}=\arcsin\left(({z_{\textrm{VA}}-z_{k}})/({\|\mathbf{x}_{\textrm{VA}}-\mathbf{v}_{k}\|})\right)$. 

\subsection{Source is a small object}

For small objects (SPs), the relations are largely a special case
of the VAs. We here only note the differences, considering an SP with
location $\mathbf{x}_{\text{SP}}$: $\tau=\Vert\mathbf{x}_{\mathrm{SP}}-\mathbf{x}_{\mathrm{BS}}\Vert/c+\Vert\mathbf{x}_{\mathrm{SP}}-\mathbf{v}_{k}\Vert/c+B$; $\phi^{\textrm{az}}=\arctan\left({y_{\textrm{SP}}}/{x_{\textrm{SP}}}\right)$
and $\phi^{\textrm{el}}=\arcsin\left(({z_{\textrm{SP}}-z_{\textrm{BS}}})/({\|\mathbf{x}_{\textrm{SP}}-\mathbf{x}_{\mathrm{BS}}\|})\right)$; and  $\theta^{\textrm{az}}=\arctan\left(({y_{\textrm{SP}}-y_{k}})/({x_{\textrm{SP}}-x_{k}})\right)-\alpha_{k}$
and $\theta^{\textrm{el}}=\arcsin\left(({z_{\textrm{SP}}-z_{k}})/({\|\mathbf{x}_{\textrm{SP}}-\mathbf{v}_{k}\|})\right)$.

\section{Proof of Expected Likelihood (\ref{eq:likelihood})}\label{sec:weightProof}

% From \eqref{eq:PHDtoPDF}, we find that under a PPP prior approximation,
From \eqref{eq:PHDtoPDF}, we find $f(\mathcal{X}|\mathbf{s}_{0:k}^{i},\mathcal{Z}_{1:k-1})\propto \prod_{(\mathbf{x},m)\in\mathcal{X}}D_{k|k-1}^{i}(\mathbf{x},m)$ under a PPP prior approximation,
 while with $\mathcal{Z}_k=\{ \mathbf{z}_1,\ldots, \mathbf{z}_{|\mathcal{Z}_k|}\}$, the function  $f(\mathcal{Z}_{k}|\mathcal{X},\mathbf{s}_{0:k}^{i},\mathcal{Z}_{1:k-1})$ can be expressed as \cite[eq.~(13)]{Garcia-Fernandez2018}
\begin{align}
f(\mathcal{Z}_{k}|\mathcal{X},\mathbf{s}_{0:k}^{i},\mathcal{Z}_{1:k-1})=
\sum_{\mathcal{U}\uplus\mathcal{X}_{1}\ldots\uplus\mathcal{X}_{|\mathcal{Z}_{k}|}=\mathcal{X}}\prod_{(\mathbf{x},m)\in\mathcal{U}}(1-p_{\text{D}}^{i}(\mathbf{x},m))\times\prod_{r=1}^{|\mathcal{Z}_{k}|}\ell(\mathbf{z}_{r}|\mathcal{X}_{r},\mathbf{s}_{k}^{i}),
\end{align}
where $\uplus$ denotes the disjoint union and \cite[eq.~(14)]{Garcia-Fernandez2018}
\begin{align}
\ell(\mathbf{z}_{r}|\mathcal{X}_{r},\mathbf{s}_{k}^{i})= \label{eq:ellfunctiondef}
\begin{cases}
p_{\text{D}}^{i}(\mathbf{x},\mathbf{s}_{k}^{i},m)f(\mathbf{z}_{r}|\mathbf{x},\mathbf{s}_{k}^{i},m) & \mathcal{X}_{r}=\{(\mathbf{x},m)\},\\
c(\mathbf{z}_{r}) & X_{r}=\emptyset,\\
0 & \text{else}.
\end{cases}
\end{align}
%Hence
%\begin{align*}
% & f(\mathcal{Z}_{k}|\mathcal{X},\mathbf{s}_{0:k}^{i},\mathcal{Z}_{1:k-1})f(\mathcal{X}|\mathbf{s}_{0:k}^{i},\mathcal{Z}_{1:k-1})\\
% & =\sum_{\mathcal{U}\uplus\mathcal{X}_{1}\ldots\uplus\mathcal{X}_{|\mathcal{Z}_{k}|}=\mathcal{X}}\prod_{(\mathbf{x},m)\in\mathcal{U}}(1-p_{\text{D}}^{i}(\mathbf{x},m))D_{k|k-1}^{i}(\mathbf{x},m)\times\prod_{r=1}^{|\mathcal{Z}_{k}|}\ell(\mathbf{z}_{r}|\mathcal{X}_{r},\mathbf{s}_{k}^{i})D_{k|k-1}^{i}(\mathcal{X}_{r}).
%\end{align*}
Introducing a PPP density
$q(\mathcal{U})\propto\prod_{(\mathbf{x},m)\in\mathcal{U}}(1-p_{\text{D}}^{i}(\mathbf{x},m))D_{k|k-1}^{i}(\mathbf{x},m)
$
and making use of the fact that
$\int\sum_{\mathcal{X}\uplus\mathcal{Y}=\mathcal{U}}f(\mathcal{X})g(\mathcal{Y})\delta\mathcal{U}=\int f(\mathcal{X})\delta\mathcal{X}\int g(\mathcal{Y})\delta\mathcal{Y},$
we immediately find that

\begin{align}
f(\mathcal{Z}_{k}|\mathbf{s}_{0:k}^{i},\mathcal{Z}_{1:k-1})&=\int f(\mathcal{Z}_{k}|\mathcal{X},\mathbf{s}_{0:k}^{i},\mathcal{Z}_{1:k-1})f(\mathcal{X}|\mathbf{s}_{0:k}^{i},\mathcal{Z}_{1:k-1})\delta\mathcal{X} \nonumber \\
 & \propto \prod_{r=1}^{|\mathcal{Z}_{k}|}\int\ell(\mathbf{z}_{r}|\mathcal{X}_{r},\mathbf{s}_{k}^{i})D_{k|k-1}^{i}(\mathcal{X}_{r})\delta\mathcal{X}_{r} \\ &=\prod_{\mathbf{z}\in\mathcal{Z}_{k}}(c(\mathbf{z})+\sum_{m}\int p_{\text{D}}^{i}(\mathbf{x},\mathbf{s}_{k}^{i},m)f(\mathbf{z}|\mathbf{x},\mathbf{s}_{k}^{i},m)D_{k|k-1}^{i}(\mathbf{x},m)\text{d}\mathbf{x})\nonumber,
\end{align}
where the last step follows from substitution of (\ref{eq:ellfunctiondef})
and the definition of set integrals \eqref{eq:setIntegral}.

\section{Implementation of Birth Process}\label{sec:GMofBirth}
% In the birth process \eqref{eq:BirthProcessGM}, the mean $\mathbf{x}^{i,j}_{b,k}(m)$ and covariance $\mathbf{P}^{i,j}_{b,k}(m)$ corresponding to measurement $\mathbf{z}_{k,l}$
Here, we introduce the detailed implementation of the birth process \eqref{eq:BirthProcessGM}.
The mean $\mathbf{x}^{i,j(\mathbf{z})}_{b,k}(m)$ and covariance $\mathbf{P}^{i,j(\mathbf{z})}_{b,k}(m)$ corresponding to each measurement $\mathbf{z}$
are inversely estimated by using sigma point principle of the CKF \cite{HaykinCKF2009}, details of which are described in Appendix \ref{sec:InverseCKF}
For propagating the cubature points, the inverse of the nonlinear function in \eqref{eq:ObservationModel2b} is required, which in general is not defined (since a vehicle state gives rise to a noise-free measurement, but a noisy measurement may not correspond to a vehicle state). Thus, the cubature points are propagated using a simple optimization method, described in  Appendix \ref{sec:IMAXestimation}. For simplification, all indices are dropped except for the source type $m$.

%As we discussed in Section \ref{sec:ObservationModel}, it is not necessary for the SP birth to be not generated.  Thus, we generate the SP birth Gaussian when $\mathbf{x}^{i,j,\text{(SP)}}_{b,k}$ is within the FoV $r_\text{FoV}$.
%However, the vehicle location $\mathbf{x}_k^i$ can not represent the true location due to the noisy measurement, it causes the problem that the SP source is not detected even the SP measurement is given.
%To solve this problem, we adjust the FoV $r_\text{FoV}$ as much as the uncertainty of both vehicle and SP location, and the uncertainty is denoted by $r_\text{UC}$.
%\HK{The parameter $r_\text{UC}$ was not modelled, and was just set to 3. Definition of $r_\text{UC}$, and relation of $r_\text{UC}$ to the birth....}

\subsection{Mean and Covariance Estimation}\label{sec:InverseCKF}
{The mean $\mathbf{x}(m)$ and covariance $\mathbf{P}(m)$ are approximated by the following steps:}
\begin{enumerate}
  \item Factorize the covariance matrix of the measurement noise (i.e., $\bm{\Sigma}^{(n)}_{k,l}$ of \eqref{eq:ObservationModel2b}) 
    \begin{align}\label{eq:Factorize}
        \bm{\Sigma} = \mathbf{G}\mathbf{G}^\mathrm{T}.
    \end{align}
  \item Evaluate the cubature point $(c=1,...,2d_\mathbf{z})$
    \begin{align}\label{eq:CubaturePoint}
        \mathbf{z}_c = \mathbf{G}\delta_{\mathbf{z},c} + \mathbf{z},
    \end{align}
    where $d_\mathbf{z} = \dim({\mathbf{z}})$ and $\delta_{\mathbf{z},c}=\sqrt{d_\mathbf{z}}[1]_{\mathbf{z},c}$. $[1]_{\mathbf{z},c}$ is defined as the $c$-th column vector of the matrix $[\mathbf{I}_{d_\mathbf{z},d_\mathbf{z}}, -\mathbf{I}_{d_\mathbf{z},d_\mathbf{z}}]\in\mathbb{R}^{d_\mathbf{z}\times 2d_\mathbf{z}}$, where $\mathbf{I}_{d_\mathbf{z},d_\mathbf{z}}\in\mathbb{R}^{d_\mathbf{z} \times d_\mathbf{z}}$ is the identity matrix.
  \item Evaluate the propagated cubature point $\mathbf{x}_c(m) \approx \mathbf{h}^{-1}(\mathbf{z}_c|\mathbf{s},m)$ with the iterative maximum-likelihood estimation (explained further in Appendix \ref{sec:IMAXestimation}).
  \item Evaluate birth mean $\mathbf{x}(m) = \frac{1}{2d_\mathbf{z}}\sum_{c=1}^{2d_\mathbf{z}}\mathbf{x}_c(m)$
     and  covariance $\mathbf{P}(m) = \frac{1}{2d_\mathbf{z}}\sum_{c=1}^{2d_\mathbf{z}}\mathbf{x}_c(m){\mathbf{x}_c{(m)}}^{\mathrm{T}} - \mathbf{x}{(m)}{\mathbf{x}{(m)}}^{\mathrm{T}}$. 
\end{enumerate}

\subsection{Simple Optimization Problem for Propagated Cubature Point}\label{sec:IMAXestimation}
% For estimating the propagated cubature point $\mathbf{x}_c{(m)}$,
For estimating the propagated cubature point $\mathbf{x}_c{(m)}$ of step 3) in Appendix \ref{sec:InverseCKF},
we formulate an optimization problem as
\begin{align}\label{eq:BirthML}
    \operatorname*{minimize}_\mathbf{x}\big(\mathbf{h}(\mathbf{x},\mathbf{s},m)-\mathbf{z}_c\big)^\mathrm{T}\bm{\Sigma}^{-1}\big(\mathbf{h}(\mathbf{x},\mathbf{s},m)-\mathbf{z}_c\big),
\end{align}
where $\mathbf{h}(\mathbf{x},\mathbf{s},m)$ is the observation function for the source type $m$ with the source location $\mathbf{x}$ and vehicle state $\mathbf{s}$, and $\mathbf{z}_c\in\mathbb{R}^{d_z}$ is the evaluated cubature point in \eqref{eq:CubaturePoint}.
Note that the used function $\mathbf{h}(\cdot)$ are determined with respect to the source type $m$, which were described in Appendix \ref{sec:Geometry}.
However, \eqref{eq:BirthML} does not admit a closed-form solution, and an optimal point is determined in a iterative manner.
% The iterative maximum-likelihood estimation is introduced at time $k$, and the number of Gaussians for the birth is equal to the number of signal paths $L_k^{(n)}+1$ at the receiver of the vehicle $n$.
% Thus, the time index $k$ is dropped.
The optimum point at the iteration $\epsilon+1$ is designed as $\bar{\mathbf{x}}_{\epsilon+1}{(m)} = (1-\eta)\bar{\mathbf{x}}_{\epsilon}{(m)} + \eta\tilde{\mathbf{x}}_{\epsilon+1}{(m)}$, 
where the design parameter $\eta$ is set to 0.2, and the initial point $\bar{\mathbf{x}}_0$ is obtained by geometric relations in Appendix \ref{sec:Geometry}.
$\tilde{\mathbf{x}}_{\epsilon+1}$ is calculated as $\tilde{\mathbf{x}}_{\epsilon+1} = \operatorname*{argmin}_\mathbf{x} {\mathbf{A}(\mathbf{x},m)}^\mathrm{T}\bm{\Sigma}^{-1}\mathbf{A}(\mathbf{x},m)$,
where $\mathbf{A}(\mathbf{x},m)$ is denoted by $\mathbf{A}(\mathbf{x},m)
        =\mathbf{H}(\bar{\mathbf{x}}_{\epsilon},\mathbf{s},m)(\mathbf{x}-\bar{\mathbf{x}}_{\epsilon}) + \mathbf{h}(\bar{\mathbf{x}}_{\epsilon},\mathbf{s},m) -\mathbf{z}_c$, 
where $\mathbf{H}(\bar{\mathbf{x}}_{\epsilon},\mathbf{s},m)\in \mathbb{R}^{5 \times 3}$ is a Jacobian matrix, and is  calculated by the finite difference method \cite{Smith1985}. The difference is set to $10^{-3}$. The iterative method is performed until the cost \eqref{eq:BirthML} increases, and then $\mathbf{x}_c$ is determined.

\section{Pseudo-code for Map Update}\label{sec:pseudoCodeMap}
The pseudo-code is provided in Algorithms \ref{alg:MMPHD_Correction}--\ref{alg:MMPHD_Correction_step2}.

\alglanguage{pseudocode}
\begin{algorithm*}[!t]
\footnotesize
\caption{\Rb{Map Correction}}\label{alg:MMPHD_Correction}
\textbf{Input}: $\{D^i_{k|k-1}(\mathbf{x},m)\}_{i=1}^I~~\forall m,$ where $D^i_{k|k-1}(\mathbf{x},m)$ $=$ $\{\gamma^{i,j}_{k|k-1}(m),\mathbf{x}^{i,j}_{k|k-1}(m),\mathbf{P}^{i,j}_{k|k-1}(m)\}_{j=1}^{J^{i,(m)}_{k|k-1}}$, $\{\mathbf{s}_{k|k-1}^i,\omega_{k|k-1}^i\}_{i=1}^I, \mathcal{Z}_k$. \\
\textbf{Output}: $\{D^i_{k|k}(\mathbf{x},m)\}_{i=1}^I~~\forall m,$ where $D^i_{k|k}(\mathbf{x},m)=\{\gamma^{i,j}_{k|k}(m),\mathbf{x}^{i,j}_{k|k}(m),\mathbf{P}^{i,j}_{k|k}(m)\}_{j=1}^{J^{i,(m)}_{k|k}}$.
\begin{algorithmic}[1]
	\For {$i=1$ to $I$} (particle sample)
	    \State Compute PHD update components by CKF (described in Algorithm \ref{alg:MMPHD_Correction_step1})
        \State PHD update (described in Algorithm \ref{alg:MMPHD_Correction_step2})
	\EndFor
\end{algorithmic}
\end{algorithm*}

\alglanguage{pseudocode}
\begin{algorithm*}[!t]
\footnotesize
\caption{\Rb{Compute PHD update components by CKF}}\label{alg:MMPHD_Correction_step1}
\begin{algorithmic}[1]
    \For {$m=\{\text{BS,VA,SP}\}$} (source type)
        \For{$j=1$ to $J^{i}_{k|k-1}(m)$} 
            \State $\mathbf{P}^{i,j}_{k|k-1}(m) = \mathbf{G}^{i,j}_k(m){\mathbf{G}^{i,j}_k(m)}^\text{T}$
            \For{$c=1$ to $2d_\mathbf{x}$}
                \State $\mathbf{X}^{i,j}_{c,k}(m) = \mathbf{x}^{i,q}_{k|k-1}(m)+\mathbf{G}^{i,j}_k(m)\delta_{\mathbf{x},c}$, $\mathbf{Z}_{c,k}^{i,j}(m) = \mathbf{h}_k(\mathbf{x}^{i,j}_{k|k-1}(m),\mathbf{s}^{i}_{k},m)$
            \EndFor
            \State ${\mathbf{Z}}^{i,j}_{k}(m) = \frac{1}{2d_\mathbf{x}}\sum_{c=1}^{2d_\mathbf{x}}\mathbf{Z}^{i,j}_{c,k}(m)$
            \State
            $\mathbf{S}^{i,j}_{\mathbf{z}\mathbf{z},k}(m)=\frac{1}{2d_\mathbf{x}}\sum_{c=1}^{2d_\mathbf{x}}\mathbf{Z}_{c,k}^{i,j}(m){\mathbf{Z}_{c,k}^{i,j}(m)}^\text{T}-{\mathbf{Z}}^{i,j}_{k}(m){{\mathbf{Z}}^{i,j}_{k}(m)}^\text{T}+\bm{\Sigma}_{k,l}$
            \If{$m=\{\text{VA,SP}\}$}
                \State ${\mathbf{X}}^{i,j}_{k}(m) = \frac{1}{2d_\mathbf{x}}\sum_{c=1}^{2d_\mathbf{x}}\mathbf{X}^{i,j}_{c,k}(m)$
                \State
                $\mathbf{S}^{i,j}_{\mathbf{x}\mathbf{z},k}(m)=\frac{1}{2d_\mathbf{x}}\sum_{c=1}^{2d_\mathbf{x}}\mathbf{X}_{c,k}^{i,j}(m){\mathbf{Z}_{c,k}^{i,j}(m)}^\text{T}-{\mathbf{X}}^{i,j}_{k}(m){{\mathbf{Z}}^{i,j}_{k}(m)}^\text{T}$
                \State
                $\mathbf{K}^{i,j}_{k}(m) = \mathbf{S}^{i,j}_{\mathbf{x}\mathbf{z},k}(m){\mathbf{S}^{i,j}_{\mathbf{z}\mathbf{z},k}(m)}^{-1}$
                \State
                $\mathbf{P}^{i,j}_{k|k}(m) = \mathbf{P}^{i,j}_{k|k-1}(m)-\mathbf{K}^{i,j}_{k}(m)\mathbf{S}^{i,j}_{\mathbf{z}\mathbf{z},k}(m){\mathbf{K}^{i,j}_{k}(m)}^\text{T}$
            \EndIf
        \EndFor
        \State $q=0$
        \For{$\mathbf{z}\in \mathcal{Z}_k$}
        \State $q = q + 1$
	        \For{$j=1$ to $J^{i}_{k|k-1}(m)$} 
	            \If{$j = j(\mathbf{z})$}
	                \State  $\mu^{i,qJ^{i}_{k|k-1}(m)+j}_k(m)=\gamma^{i,j}_{k|k-1}(m)$
	            \Else~($j \neq j(\mathbf{z})$)
	                \State $\mu^{i,qJ^{i}_{k|k-1}(m)+j}_k(m)=p^{i,j}_{D,k}(m)\gamma^{i,j}_{k|k-1}(m)\mathcal{N}(\mathbf{z};\mathbf{h}_k(\mathbf{x}^{i,j}_{k|k-1}(m),\mathbf{s}_k^i,m),\mathbf{S}_{\mathbf{z}\mathbf{z},k}^{i,j}(m))$
	            \EndIf
	        \EndFor
        \EndFor
    \EndFor
\end{algorithmic}
\end{algorithm*}

\clearpage

\alglanguage{pseudocode}
\begin{algorithm*}[!t]
\footnotesize
\caption{\Rb{PHD update}}\label{alg:MMPHD_Correction_step2}
\begin{algorithmic}[t]
    \For {$m=\{\text{VA,SP}\}$} (source type)
        \For{$j=1$ to $J^{i}_{k|k-1}(m)$} (missed detections)
            \If{$j$ is a birth}
                \State $\gamma^{i,j}_{k|k}(m) = 0$, $\mathbf{x}^{i,j}_{k|k}(m) = \mathbf{x}^{i,j}_{k|k-1}(m)$, $\mathbf{P}^{i,j}_{k|k}(m) = \mathbf{P}^{i,j}_{k|k-1}(m)$
            \Else~($j$ is not a birth)
                \State $\gamma^{i,j}_{k|k}(m) = \{1-p^{i,j}_{D,k}(m)\}\gamma^{i,j}_{k|k-1}(m)$, $\mathbf{x}^{i,j}_{k|k}(m) = \mathbf{x}^{i,j}_{k|k-1}(m)$, $\mathbf{P}^{i,j}_{k|k}(m) = \mathbf{P}^{i,j}_{k|k-1}(m)$
            \EndIf
        \EndFor
        \State $q = 0$;
        \For{$\mathbf{z}\in \mathcal{Z}_k$} (detections)
        \State $q = q + 1$
	        \For{$j=1$ to $J^{i}_{k|k-1}(m)$} 
	            \If{$j = j(\mathbf{z})$}
	                \State $\mathbf{x}^{i,qJ^{i}_{k|k-1}(m)+j}_{k|k}(m) = \mathbf{x}^{i,j}_{k|k-1}(m)$, $\mathbf{P}^{i,qJ^{i}_{k|k-1}(m)+j}_{k|k}(m) = \mathbf{P}^{i,j}_{k|k-1}(m)$
	            \Else~($j \neq j(\mathbf{z})$)
	                \State $\mathbf{x}^{i,qJ^{i}_{k|k-1}(m)+j}_{k|k}(m) = \mathbf{x}^{i,j}_{k|k-1}(m) + \mathbf{K}^{i,j}_{k}(m)(\mathbf{z}-{\mathbf{Z}_k}^{i,j}(m))$
	                \State
	                $\mathbf{P}^{i,qJ^{i}_{k|k-1}(m)+j}_{k|k}(m) = \mathbf{P}^{i,j}_{k|k}(m)$
	            \EndIf
	            \State  $\gamma^{i,qJ^{i}_{k|k-1}(m)+j}_{k|k}(m)= \mu^{i,qJ^{i}_{k|k-1}(m)+j}_k(m)/(c(\mathbf{z})+ \sum_{m'=\{\text{BS,VA,SP}\}}\sum_{j'=1}^{J^{i}_{k|k-1}(m')}$\par
	            ~~~\hskip\algorithmicindent$\mu^{i,qJ^{i}_{k|k-1}(m')+j'}_k(m'))$
	        \EndFor
        \EndFor
        \State $J^{i}_{k|k}(m)=(q+1)J^{i}_{k|k-1}(m)$ 
    \EndFor
    \State $\mathbf{x}^{i,j}_{k|k}(\text{BS}) = \mathbf{x}^{i,j}_{k|k-1}(\text{BS}), \mathbf{P}^{i,j}_{k|k}(\text{BS}) = \mathbf{P}^{i,j}_{k|k-1}(\text{BS}), \gamma^{i,j}_{k|k}(\text{BS})=1, J^{i}_{k|k}(\text{BS}) = 1$
\end{algorithmic}
\end{algorithm*}

\bibliographystyle{IEEEtran}
\bibliography{bibliography}

\end{document}